\documentclass{emulateapj}

\slugcomment{Received 2005 November 19; accepted 2006 April 5}

\shorttitle{Ly$\alpha$ emitters at $z=6.5$}
\shortauthors{Kashikawa et al.}

\begin{document}


\title{The End of the Reionization Epoch Probed by Ly$\alpha$ Emitters at $z=6.5$ \\
in the Subaru Deep Field\altaffilmark{1,2}}


\author{
Nobunari Kashikawa\altaffilmark{3,4}, 
Kazuhiro Shimasaku\altaffilmark{5,6}, 
Matthew A. Malkan\altaffilmark{7}, 
Mamoru Doi\altaffilmark{8}, 
Yuichi Matsuda\altaffilmark{9},\\
Masami Ouchi\altaffilmark{10}, 
Yoshiaki Taniguchi\altaffilmark{11},
Chun Ly\altaffilmark{7},
Tohru Nagao\altaffilmark{3,12}, 
Masanori Iye\altaffilmark{3,4},\\ 
Kentaro Motohara\altaffilmark{8}, 
Takashi Murayama\altaffilmark{11}, 
Kouji Murozono\altaffilmark{5}, 
Kyoji Nariai\altaffilmark{13}, 
Kouji Ohta\altaffilmark{9}, \\
Sadanori Okamura\altaffilmark{5,6}, 
Toshiyuki Sasaki\altaffilmark{14}, 
Yasuhiro Shioya\altaffilmark{11}, and
Masayuki Umemura\altaffilmark{15}
}




\email{kashik@zone.mtk.nao.ac.jp}


\altaffiltext{1}{The data presented herein were partly obtained at the W. M. Keck Observatory, which is operated as a scientific partnership among the California Institute of Technology, the University of California, and the National Aeronautics and Space Administration. 
The Observatory was made possible by the generous financial support of the W. M. Keck Foundation.}
\altaffiltext{2}{Based in part on data collected at the Subaru Telescope, which is operated by the National Astronomical Observatory of Japan.}
\altaffiltext{3}{Optical and Infrared Astronomy Division, National Astronomical Observatory, Mitaka, Tokyo 181-8588, Japan.}
\altaffiltext{4}{Department of Astronomy, School of Science, Graduate University for Advanced Studies, Mitaka, Tokyo 181-8588, Japan.}
\altaffiltext{5}{Department of Astronomy, University of Tokyo, Hongo, Tokyo 113-0033, Japan.}
\altaffiltext{6}{Research Center for the Early Universe, University of Tokyo, Hongo, Tokyo 113-0033, Japan.}
\altaffiltext{7}{Department of Physics and Astronomy, University of California, Los Angeles, CA 90095-1547.}
\altaffiltext{8}{Institute of Astronomy, University of Tokyo, Mitaka, Tokyo 181-8588, Japan.}
\altaffiltext{9}{Department of Astronomy, Graduate School of Science, Kyoto University, Kyoto 606-8502, Japan.}
\altaffiltext{10}{Space Telescope Science Institute, 3700 San Martin Drive, Baltimore, MD 21218.}
\altaffiltext{11}{Astronomical Institute, Graduate School of Science, Tohoku University, Aramaki, Aoba, Sendai 980-8578, Japan.}
\altaffiltext{12}{INAF --- Osservatorio Astrofisico di Arcetri, Largo Enrico Fermi 5, 50125 Florence, Italy.}
\altaffiltext{13}{Department of Physics, Meisei University, 2-1-1 Hodokubo, Hino, Tokyo 191-8506, Japan.}
\altaffiltext{14}{Subaru Telescope, National Astronomical Observatory of Japan, 650 North A'ohoku Place, Hilo, HI 96720.}
\altaffiltext{15}{Center for Computational Physics, University of Tsukuba, 1-1-1 Tennodai, Tsukuba 305-8571, Japan.}
\setcounter{footnote}{15}



\begin{abstract}
We report an extensive search for Ly$\alpha$ emitters (LAEs) at $z=6.5$ in the Subaru Deep Field.
Subsequent spectroscopy with Subaru and Keck identified eight more LAEs, giving a total of $17$ spectroscopically confirmed LAEs at $z=6.5$.
Based on this spectroscopic sample of $17$, complemented by a photometric sample of $58$ LAEs, we have derived a more accurate Ly$\alpha$ luminosity function of LAEs at $z=6.5$, which reveals an apparent deficit at the bright end of $\sim0.75$ mag fainter $L^*$, compared with that observed at $z=5.7$.
The difference in the LAE luminosity functions between $z=5.7$ and $6.5$ is significant at the $3$ $\sigma$ level, which is reduced to $2$ $\sigma$ when cosmic variance is taken into account.
This result may imply that the reionization of the universe has not been completed at $z=6.5$.
We found that the spatial distribution of LAEs at $z=6.5$ was homogeneous over the field.
We discuss the implications of these results for the reionization of the universe.
\end{abstract}






\section{Introduction}

The cosmic reionization was undoubtedly one of the major turning points in the early universe.
The measurement of cosmic microwave background (CMB) temperature polarization by the {\it Wilkinson Microwave Anisotropy Probe} ({\it WMAP}) implies an early reionization at $z=10.9^{+2.7}_{-2.3}$ \citep{pag06}, and the complete Gunn-Peterson trough of Sloan Digital Sky Survey (SDSS) QSOs suggests that cosmic reionization ended at $z\sim 6$ \citep{fan02}.
There are disputes over when and how the reionization has taken place, and which objects were responsible for it.
Although QSOs are expected to be the main contributor of ionizing photons at the bright end of the luminosity function (LF) of ionizing sources, the QSO population alone cannot account for all the required ionizing photons \citep{wil05}, and star-forming galaxies like Lyman break galaxies (LBGs) and Ly$\alpha$ emitters (LAEs) at the reionization epoch are the only alternatives that could dominate at the faint end.
The census of observable galaxies at this epoch is sensitive to the ionization fraction of the universe \citep{yan04, mal04, bou05, sti05, bun06}.
It is expected that the surrounding neutral intergalactic medium (IGM) attenuates the Ly$\alpha$ photons so significantly that the number density decline of LAEs provides a useful observational constraint on the reionization epoch \citep{hai99, rho01, hu02}.

\begin{deluxetable*}{llllllllll}
\tabletypesize{\footnotesize}
\tablecaption{Summary of Spectroscopic Identifications\label{tab_sum}}
\tablewidth{0pt}
\tablehead{
\colhead{Observational run} & \colhead{Instrument} & \colhead{$N_{tot}$\tablenotemark{a}} & \colhead{$N_{cand}$\tablenotemark{b}} & \colhead{Ly$\alpha$} & \colhead{H$\alpha$} & \colhead{$[$O {\sc iii}$]$} & \colhead{$[$O {\sc ii}$]$} & \colhead{Single Line\tablenotemark{c}} & \colhead{Ref.}
}
\startdata
2002-2003 & Subaru FOCAS  & 20 & 13 & 10(9)\tablenotemark{d} & 0 & 4\tablenotemark{e}  & 0 & 3(5)\tablenotemark{d}+3\tablenotemark{e} & 1,2\\
2004      & Keck II DEIMOS & 14 & 6 & 5                      & 1 & 4  & 4 & 0                     & 3\\
2004      & Subaru FOCAS  & 19 & 3 & 2                      & 1 & 14 & 0 & 2                     & 3
\enddata
\tablenotetext{a}{The total number of objects for which we obtained a spectroscopic signal.}
\tablenotetext{b}{The total number of z6p5LAE candidates that meet our photometric selection criteria.}
\tablenotetext{c}{These possess neither line asymmetry with large $S_w$ as in a LAE nor doublet features as in an $[$O {\sc ii}$]$ emitter.}
\tablenotetext{d}{Although nine LAEs and five single-line objects were reported in T05, one of the single-line objects, SDF J132520.4+273459, was reobserved on the 2004 Keck II DEIMOS run and was found to be a LAE.
Another single-line object, SDF J132518.4+272122 was concluded to be a LAE in this study based on the $S_w$ classification.}
\tablenotetext{e}{The spectra of four $[$O {\sc iii}$]$ emitters and three single-line objects were obtained as for the $NB921$ strong emitter on the $2002$ run.\\
References.--- (1) Kodaira et al. 2003; (2) Taniguchi et al. 2005; (3) this paper.}
\end{deluxetable*}

There have been great advances over the past three years in detecting distant galaxies at the edge of the cosmic reionization era beyond $z=6$ in both dropout searches \citep{dic04, bou03, bou05, kne04, pel04, sta05} to find their strong Lyman breaks and narrowband (NB) searches \citep{hu02, cub03, aji03, kod03, rho04, mal04, tan05} to detect their Ly$\alpha$ emission lines.
Complementary to these photometric surveys are direct spectroscopic approaches based on slitless spectroscopic searches \citep{kur04, mal05} and blind slit searches \citep{tra04, mar06}.
The Subaru Telescope plays an important role in these challenging searches for high-$z$ populations, in particular, in the Subaru Deep Field (SDF).
The major goal of this project is to construct large samples of LBGs at $z\simeq 4-5$ and of LAEs at $z \simeq 4.8$, $5.7$, and $6.6$, and to make detailed studies of these very high $z$ galaxy populations.
The SDF's wide-field imaging increases the chance of discovering rare objects, such as the most distant galaxies.
In addition to the improved detectability, the wide field of view is less sensitive to the potentially large cosmic scatter in the reionization history \citep{bar04, som04}.
Following our first discovery of a couple of LAEs at $z=6.5$ (z6p5LAEs; \citealp{kod03}), \citet[T05]{tan05} have revealed for the first time a statistically useful sample of nine spectroscopically identified z6p5LAEs and estimated their total amount of star formation rate density at this high-$z$ end.
\citet{nag04, nag05} also serendipitously discovered strong Ly$\alpha$ emission at $z>6$ from an $i'$-drop-selected sample in the SDF.
Our SDF LAE sample was obtained from a general blank field without resorting to amplification of gravitational lensing by foreground clusters, providing reliable statistics about their number density, LF, and cosmic star formation rate density.
High-$z$ surveys using gravitational lensing are complementary to our survey, because they detect low-luminosity sources \citep{ell01, hu02, san04, kne04}.

In this paper, we report the discovery of eight additional spectroscopically identified z6p5LAEs, which enables a more accurate estimation of their LF.
The LF beyond $z=6$ puts a critical constraint on the reionization epoch, as well as on the ionizing photon budget.
The Ly$\alpha$ photons are absorbed when passing through the neutral IGM; therefore, it is naturally expected that the LF of LAEs should decline as it traces earlier times in the reionization epoch.
Consequently, the observed abundance of LAEs during the reionization period should indicate the neutral fraction of IGM hydrogen $x_{\rm H I}^{\rm IGM}$ \citep{mir98}.
\citet{mal04} and \citet{ste05} found no significant evolution of the LF between $z=5.7$ and $6.5$, implying that the neutral fraction of the universe is already low at $z=6.5$.
However, their LF estimate at $z=6.5$ was poorly determined, since it was combined from several independent data sets with different selection criteria.
On the other hand, there are updated model predictions for the LAE's LF during the reionization epoch \citep{hai99,del05,hai05}.

In addition, we evaluate the inhomogeneity of the sky distribution for our z6p5LAE sample.
The high-$z$ galaxy survey in a general field also has an advantage for determining spatial clustering.
The $i$-dropout method generally samples a wide redshift range at $5.7<z<6.2$, which corresponds to a comoving distance as deep as $\sim200$ Mpc along the line of sight.
The large-scale structure within this large volume probed by the $i$-dropout method would be diluted in sky projection and thus cannot be revealed unless large spectroscopic samples are obtained.
On the other hand, NB searches exploring only a small redshift range are more sensitive to the large-scale structure, although their thin slices of the universe are, at the same time, likely to be affected by cosmic variance \citep{shi04}.
The detection of an inhomogeneous distribution of galaxies beyond $z=6$ would be of great interest not only because it would reveal the primeval large-scale structure \citep{sti05, mal05, ouc05}, but also because it could provide evidence of patchy reionization, in which Ly$\alpha$ flux is attenuated in one field and not in the other.
In the reionization epoch, ionizing sources like LAEs would make cosmological H {\sc ii} regions around them \citep{mir00}.
The ionized spheres around adjacent LAEs will overlap, and the space overdensity of these ionizing sources would form a large enough H {\sc ii} region around them to allow high transmission of their Ly$\alpha$ photons prior to reionization \citep{wyi05, fur06}.
The maximum extent of the overlapped ionized regions is predicted to be $\sim10$ physical Mpc \citep{wyi04, fur05}, which is comparable to the field of view of the SDF.
A measurement of inhomogeneity in the spatial distribution of such a high-$z$ population would draw a much more precise picture of the reionization process than has been previously possible.

This paper is organized as follows.
In \S~2, we describe our new spectroscopically identified z6p5LAE sample.
In \S~3, we derive the Ly$\alpha$ LF of our z6p5LAE sample making comparison with LFs at $z=5.7$.
An estimate of inhomogeneity in the sky distribution for our z6p5LAE sample is presented in \S~4.
We present the composite spectrum of our spectroscopically confirmed z6p5LAE sample in \S~5.
Some discussions of the implications for reionization based on our results are made in \S~6, and the summary of the paper is given in \S~7.

Throughout the paper, we analyze in the flat $\Lambda$CDM model: $\Omega_m=0.3$, $\Omega_\Lambda=0.7$, and $H_0=70$ $h_{70}$kms$^{-1}$ Mpc$^{-1}$. 
These parameters are consistent with recent CMB constraints \citep{spe06}.
Magnitudes are given in the AB system.

\section{New Spectroscopic Confirmation}

\epsscale{1.5}
\begin{figure}
\hspace*{-1.0cm}
\plotone{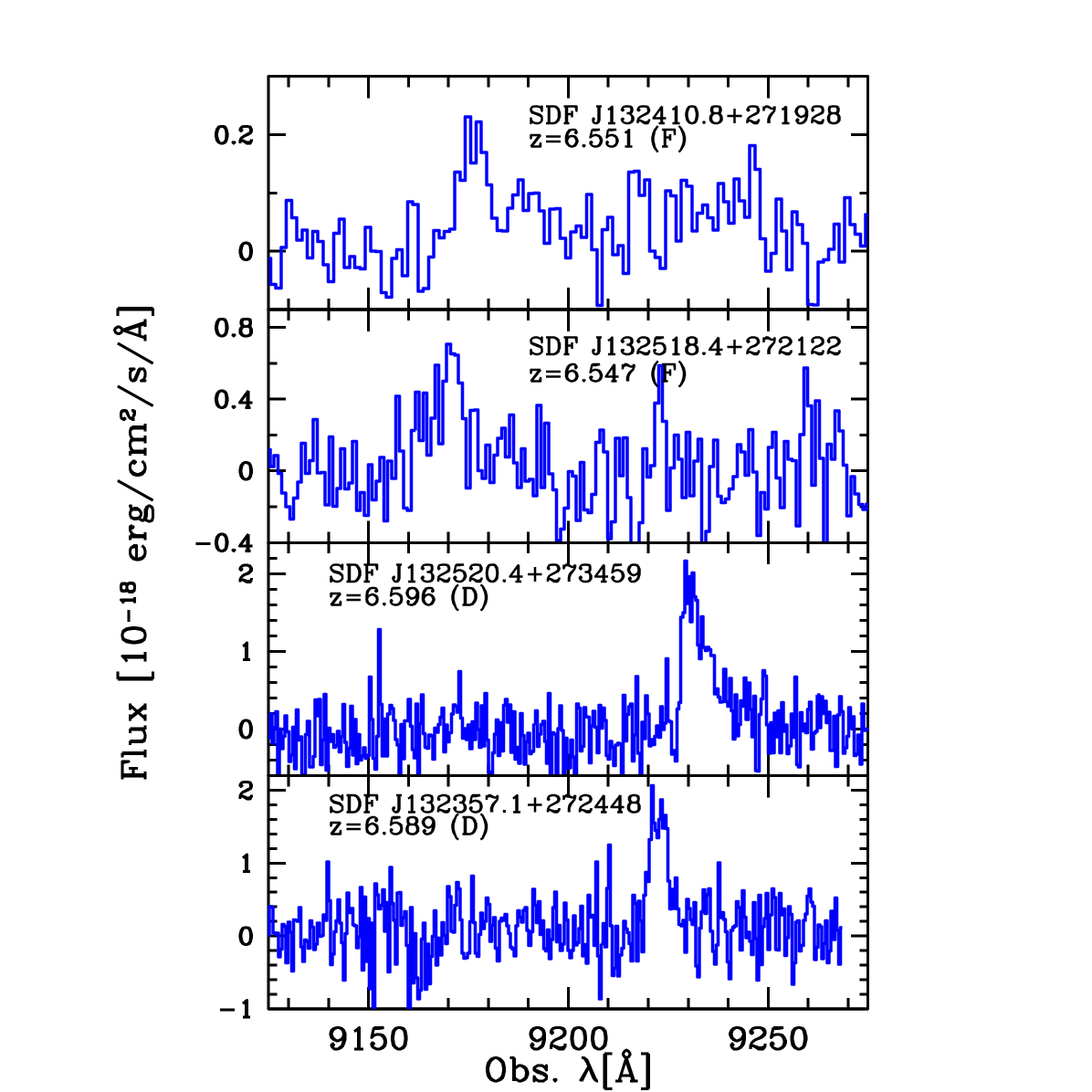}
\hspace*{-1.0cm}
\plotone{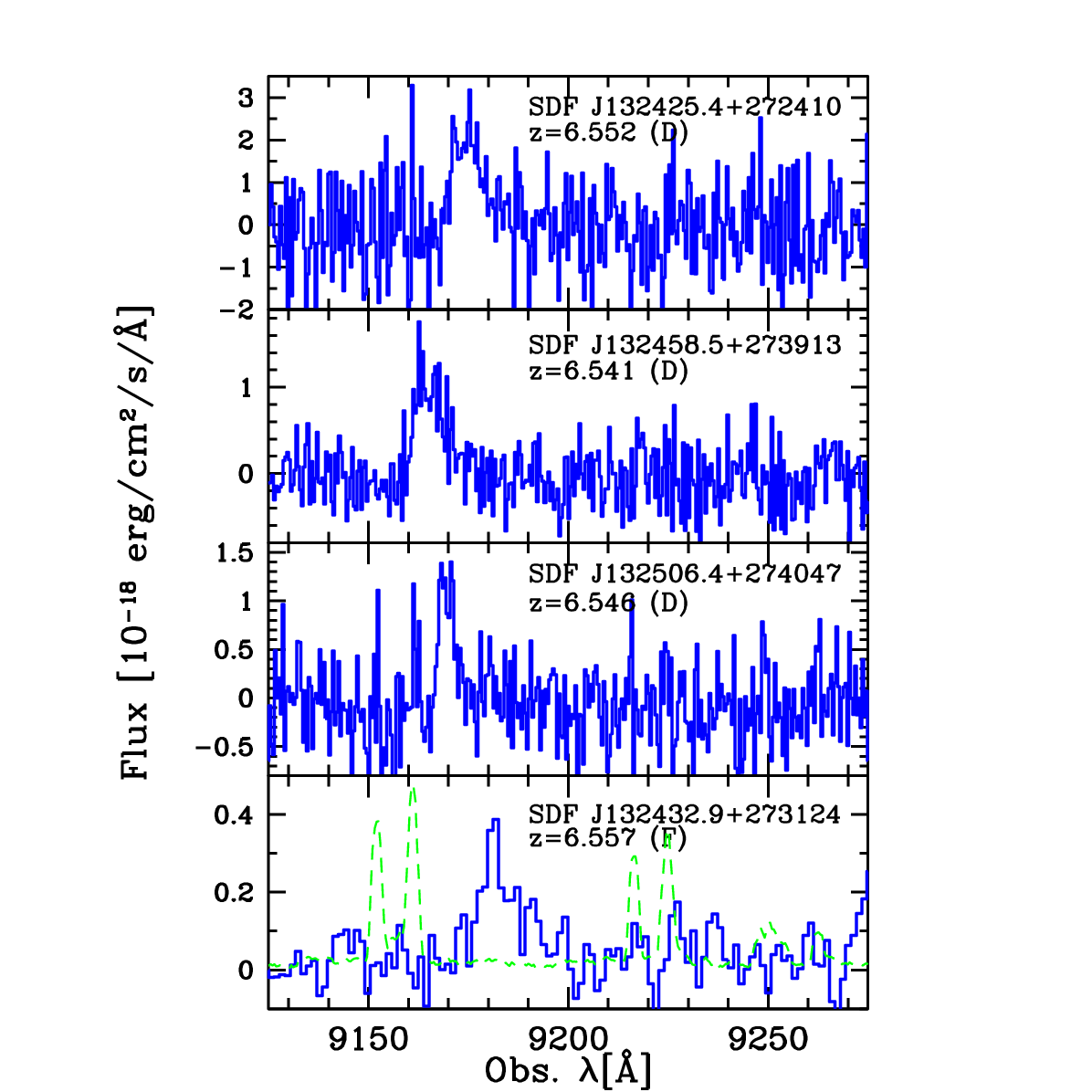}
\epsscale{1.0}
\caption{
Spectra of eight spectroscopically confirmed z6p5LAEs.
^^ ^^ F" (^^ ^^ D") in parentheses indicates that the object was observed with FOCAS (DEIMOS).
The sky spectrum is overplotted on the bottom panel with an arbitrary flux scale.
The spectrum of SDF J132518.4+272122 which was identified as a LAE in this study already appeared in T05.
\label{fig_spe}}
\end{figure}

Our z6p5LAE photometric candidate sample in the SDF was presented in T05, in which sample selection and star formation rate density were discussed.
The sample was based on the flux excess objects in narrowband $NB921$ ($\lambda_c=9196$ \AA, FWHM=$132$ \AA) image compared with the very deep broadband images of the SDF \citep{kas04}.
Our comoving survey volume was as large as $2.17\times10^5$ $h_{70}^{-3}$ Mpc$^3$.
In T05, we found $58$ photometric candidates of z6p5LAEs down to $NB921=26.0$ ($5$ $\sigma$) in the effective survey region of $876$ arcmin$^2$; nine of them had been confirmed as real by spectroscopy.
In this section, we describe our extended spectroscopic confirmations of z6p5LAEs after T05.
Table~\ref{tab_sum} summarizes our spectroscopic identifications of $NB921$-excess objects in the SDF over the last three years.
In summary, we have hitherto taken spectroscopy for $22$ objects that meet the photometric selection criteria of z6p5LAE and confirmed that $16$ are really LAEs based on their asymmetric line profiles, one is an $[$O {\sc iii}$]$ emitter, and five are faint single-line emitters.
We have also included another spectroscopically identified z6p5LAE discovered serendipitously.

\begin{deluxetable*}{llccccrcrc}
\tabletypesize{\footnotesize}
\tablecaption{Spectroscopic Properties of $z=6.5$ LAEs \label{tab_laespec}}
\tablewidth{0pt}
\tablehead{
\colhead{ID\tablenotemark{a}} & \colhead{NAME} & \colhead{z\tablenotemark{b}} & \colhead{$f^{\rm spec}$(Ly$\alpha$)\tablenotemark{c}} & \colhead{$L^{\rm spec}$(Ly$\alpha$)\tablenotemark{d}} & \colhead{SFR$^{\rm spec}$(Ly$\alpha$)\tablenotemark{d,e}} & \multicolumn{2}{c}{FWHM\tablenotemark{f}} & \colhead{$S_w$} & \colhead{OBS.\tablenotemark{g}}\\
   & & & ($10^{-18}$ ergs/s/cm$^2$) & ($10^{42} h_{70}^{-2}$ ergs/s) & ($M_\odot$/yr) & (\AA) & (km/s) & \multicolumn{1}{c}{(\AA)} & 
}
\startdata

11 &	SDF J132410.8+271928 &	6.551 & 5.39 & 2.64 &	2.40	& 14.6	& 477 &  $7.05 \pm  4.79$  & F\\
16 &	SDF J132518.4+272122 &	6.547 & 11.2 & 5.47 &	4.98	& 11.0	& 360 &  $4.75 \pm  2.35$  & F\\
18 &	SDF J132520.4+273459 &	6.596 & 10.0 & 4.97 &	4.52	&  8.2	& 266 & $11.58 \pm  0.54$  & D\\
30 &	SDF J132357.1+272448 &	6.589 & 10.6 & 5.25 &	4.78	&  5.9	& 192 &  $3.12 \pm  0.30$  & D\\
37 &	SDF J132425.4+272410 &	6.552 & 14.5 & 7.09 &	6.45	&  8.9	& 291 & $16.34 \pm  0.44$  & D\\
52 &	SDF J132458.5+273913 &	6.541 & 7.67 & 3.74 &	3.40	&  9.7	& 318 &  $7.63 \pm  0.71$  & D\\
55 &	SDF J132506.4+274047 &	6.546 & 9.84 & 4.80 &	4.36	&  6.3	& 206 &  $3.77 \pm  0.44$  & D\\
59 &	SDF J132432.9+273124 &	6.557 & 4.50 & 2.20 &	2.00	& 11.0	& 359 &  $4.22 \pm  2.50$  & F

\enddata
\tablenotetext{a}{The object IDs are those of T05, except ID=59, which is not listed in the photometric catalog of T05.}
\tablenotetext{b}{The redshift was derived from the wavelength of the flux peak in an observed spectrum assuming the rest wavelength of Ly$\alpha$ to be $1215$ \AA.
These measurements could be overestimated in the case of significant damping wings by IGM.
Also, the observed peak position was slightly shifted redward due to instrumental resolution. See Fig. \ref{fig_prof}.}
\tablenotetext{c}{The observed line flux corresponds to the total amount of the flux within the line profile.}
\tablenotetext{d}{No dust absorption correction was applied.}
\tablenotetext{e}{Estimated from the observed luminosities with the relation $SFR($Ly$\alpha)=9.1\times10^{-43}L($Ly$\alpha$) M$_\odot$ yr$^{-1}$ as in T05.}
\tablenotetext{f}{Corrected for instrumental broadening.}
\tablenotetext{g}{Observed with FOCAS (F) or DEIMOS (D).}

\end{deluxetable*}

\subsection{Keck II DEIMOS Spectroscopy}

The z6p5LAE candidates were observed with the Keck II DEIMOS \citep{fab03} spectrograph on UT $2004$ April $23-24$.
We also allocated slits for $NB921$-strong ($z'-NB921>1$) emitters, irrespective of their ($i'-z'$) color as a LAE criterion in order to see how our selection criteria work.
We used four multiobject spectroscopic (MOS) masks with an $830$ line mm$^{-1}$ grating and a GG495 order-cut filter for each $7000$-$9000$ s. integration time.
The central wavelength was set to $7500$ \AA~for one of the four MOS masks and $8100$ \AA~for the other three masks.
The slit width was $1\arcsec.0$ with $0.47$ \AA~pixel$^{-1}$, giving a resolving power of $\sim3600$.
The wavelength coverage was $\sim5000-10,000$ \AA, depending on position in the mask.
The typical seeing size was $0\arcsec.55-1\arcsec.0$ during the observation.
Our z6p5LAEs were almost spatially unresolved on an $NB921$ image with $0\arcsec.98$ seeing size.
Assuming that our LAEs were also spatially unresolved on the slits, the effective spectral resolution may be better, depending on the source size \citep{rho03}.
We also obtained spectra of standard stars BD +28 4211 and Feige 110 for flux calibration.
The data were reduced with the spec2d pipeline\footnote{The data reduction pipeline was developed at University of California, Berkeley, with support from National Science Foundation grant AST 00-71048.} for DEEP2 DEIMOS data reduction.

We allocated slits for $18$ target z6p5LAE candidates, as well as $NB921$-strong emitters.
Four of them were apparent $[$O {\sc iii}$]$ emitters showing their characteristic double lines and sometimes also H$\beta$, one was an H$\alpha$ emitter that has corresponding $[$O {\sc iii}$]$ emission, four were $[$O {\sc ii}$]$ doublets, five have apparently asymmetric single lines, and we did not obtain any signal from four targets.

\subsection{Subaru FOCAS Spectroscopy}

Three z6p5LAE candidates were also observed with the Subaru Faint Object Camera and Spectrograph (FOCAS) \citep{kas02} spectrograph in the MOS mode on UT $2004$ April $24-27$.
Although our primary targets for this observation were $z=5.7$ LAEs \citep{shi06} in the SDF, three slits were allocated for the z6p5LAE sample.
We also allocated slits for strong $NB921$ emitters.
The spectroscopy was made with a $300$ line mm$^{-1}$ grating and an O58 order-cut filter.
The spectra cover $5400-10,000$ \AA, with a pixel resolution of $1.34$ \AA.
The $0\arcsec.6$ wide slit gave a spectroscopic resolution of $7.1$ \AA~at $9200$ \AA~($R\sim1300$).
The spatial resolution was $0\arcsec.3$ pixel$^{-1}$ with $3$ pixel on-chip binning.
The integration time was $12,000$-$16,800$ s.
The sky conditions were fairly good with a seeing of $0\arcsec.4$-$0\arcsec.8$.
The data were reduced in a standard manner.
We also obtained spectra of standard stars Hz 44 and Feige 34 for flux calibration.

We allocated slits for $21$ targets: $14$ of them were apparent $[$O {\sc iii}$]$ emitters showing their double features and in some cases, H$\beta$; one was an H$\alpha$ emitter that has corresponding $[$O {\sc iii}$]$ emission; two had apparently asymmetric single lines; two had symmetric lines; and we did not obtain any signal for the two remaining targets.

\subsection{Spectroscopic Results}

We combined our spectroscopic sample with those reported in T05, in which nine LAEs and five single emitters are contained.
One object, SDF J132520.4+273459, which was classified as a single emitter in T05, was reobserved with DEIMOS and was found to be a LAE based on the resulting better quality spectrum.
We also obtained spectra for four $[$O {\sc iii}$]$ emitters and three single emitters classified as $NB921$ strong emitters in the same MOS observation, although these objects do not satisfy the LAE criteria and were not reported in T05.
The total spectroscopic sample for this study comprises $53$ objects.

It is difficult to identify a LAE at very high $z$ with little continuum flux and a tiny signature of Ly$\alpha$ emission.
The asymmetric line profile is the best diagnostic of high-$z$ Ly$\alpha$ emission, which results from absorption by neutral hydrogen; therefore, it strongly depends on the ionization structure in and around the high-$z$ objects.
Although some bright LAEs show the continuum breaks \citep{kod03} caused by IGM attenuation, most are too faint to detect the break at the highest $z$ epoch.
We have no other spectral features but asymmetric emission profiles that can distinguish high-$z$ LAEs from foreground $[$O {\sc ii}$]$, $[$O {\sc iii}$]$, or H$\alpha$ emitters.
To quantify this asymmetry accurately, we introduced an asymmetry statistics {\it skewness} $S$ and {\it weighted skewness} $S_w$.
Here we regarded the observed spectrum, which is basically a two-dimensional array of the flux ($f_i$) and the pixel coordinate ($x_i$), as a distribution function with an array size of $n$.
The $S$ statistic is defined as

\begin{eqnarray}
S=\frac{1}{I\sigma^3}\sum^n_i(x_i-\overline{x})^3 f_i,
\end{eqnarray}

where $I=\sum^n_i f_i$, and $\overline{x}$, $\sigma$ are the average and dispersion of $x_i$, respectively.
The $S$ indicator has an advantage of being independent of the line-profile modeling or fitting procedure.
Our statistic $S$ has a good correlation with other asymmetry indicators, $a_\lambda$ or $a_f$ \citep{rho03}, as shown in the Appendix.

The {\it weighted skewness} $S_w$ is the revised indicator of $S$ so as to be more sensitive to an asymmetry; however, it does depend on the fitting procedure.
We define the {\it weighted skewness} $S_w$ as

\begin{eqnarray}
S_w=S(\lambda_{10,r}-\lambda_{10,b}), 
\end{eqnarray}

where $\lambda_{10,r}$ and $\lambda_{10,b}$ are the wavelengths where the flux drops to $10\%$ of its peak value on the red and blue sides of the Ly$\alpha$ emission, respectively.

\epsscale{1.45}
\begin{figure}
\hspace*{-1.0cm}
\plotone{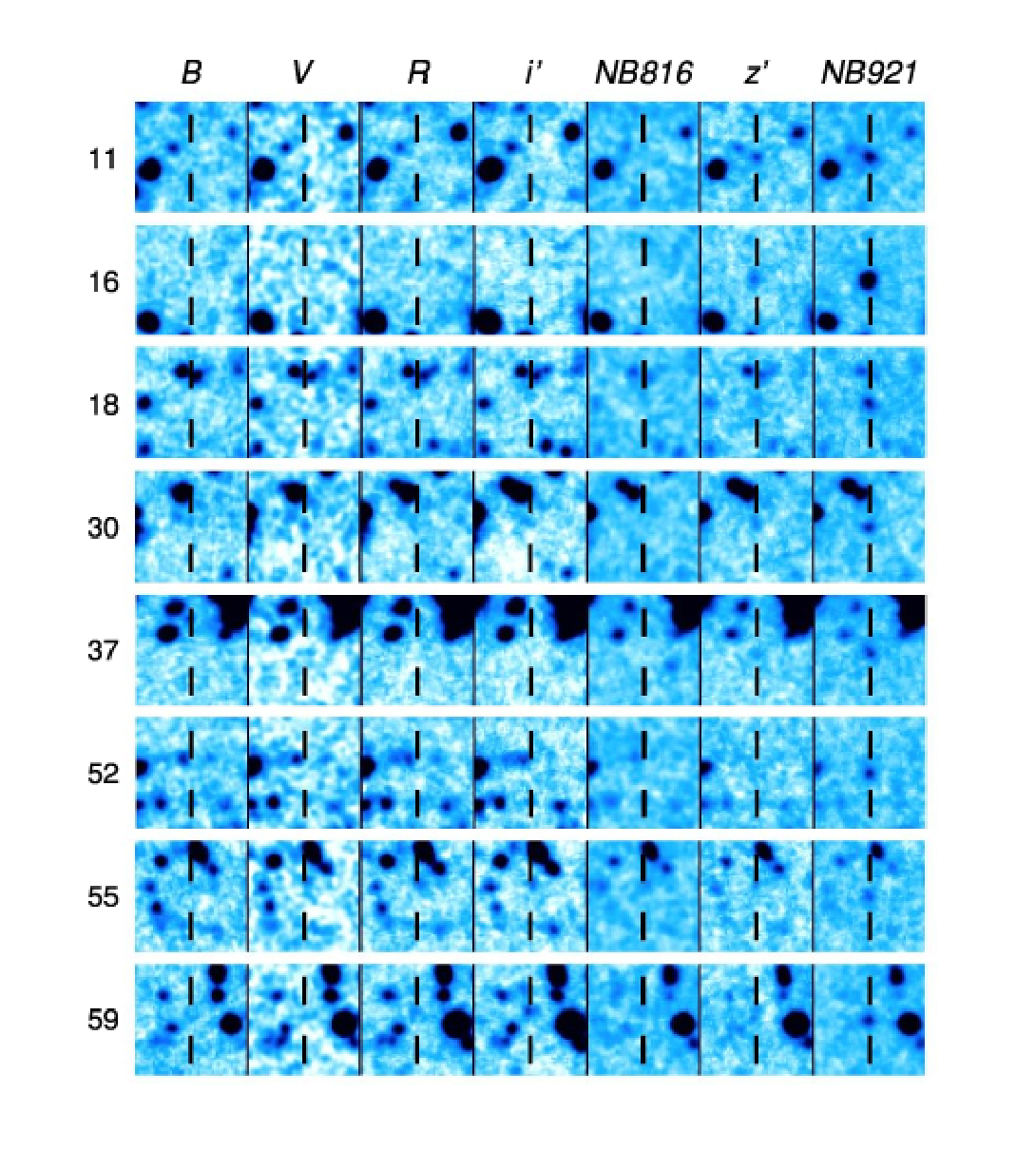}
\epsscale{1.0}
\caption{
Thumbnail images of our eight identified LAEs.
The object IDs are those of T05, except ID=59 which is not listed in the photometric catalog of T05.
The $B$-, $V$-, $R$-, $i'$-, $NB816$-, $z'$-, and $NB921$-band images are shown from left to right.
Each image is $10\arcsec$ on a side.
North is up and east is left.
\label{fig_thum}}
\end{figure}

In this study, we classified our observed emission lines based on the $S_w$ indicator as shown in the Appendix.
Table~\ref{tab_sum} summarizes the identifications for our $53$ spectra.
We have identified apparent foreground emitters, including two H$\alpha$ emitters, $22$ $[$O {\sc iii}$]$ emitters, and four $[$O {\sc ii}$]$ emitters, by their multiple emission-line signatures.
The properties of these emission-line galaxies at $z<1.2$ will be presented in our forthcoming paper \citep{ly06}.
The $S_w$ value of these apparent foreground emitters never exceeds $S_w=3$, which we set as the critical $S_w$ value to distinguish LAEs from foreground emitters.
This critical value is the same as for our $z=5.7$ LAE sample \citep{shi06}.
As a result, we have so far obtained $17$ LAEs at $z=6.5$.
All of the nine LAEs identified in T05 were classified as LAEs according to the $S_w$ criterion, and we obtained eight additional spectroscopic confirmations of LAEs in this study.\footnote{SDF J132518.4+272122 was classified as a single emitter in T05, although it shows a very red color ($i'-z'>2.21$) and high enough $S$ (0.173) and $S_w$ ($4.75$) values. 
We therefore classify it as a LAE in this study.}
The spectra of newly identified LAEs in this study are shown in Figure~\ref{fig_spe}, and their spectroscopic properties are summarized in Table~\ref{tab_laespec}.
For all eight LAE spectra, we did not detect any UV continuum fluxes significant enough to measure their equivalent widths reliably.
Nor did we detect N$_{\rm V}$ $\lambda1240$, the only accessible strong high-ionization metal line indicative of AGN activity.
We discuss the composite spectrum in \S~5.
Figure~\ref{fig_thum} presents postage stamp images of these eight LAEs in all seven bands.

\epsscale{1.4}
\begin{figure}
\hspace*{-1.0cm}
\plotone{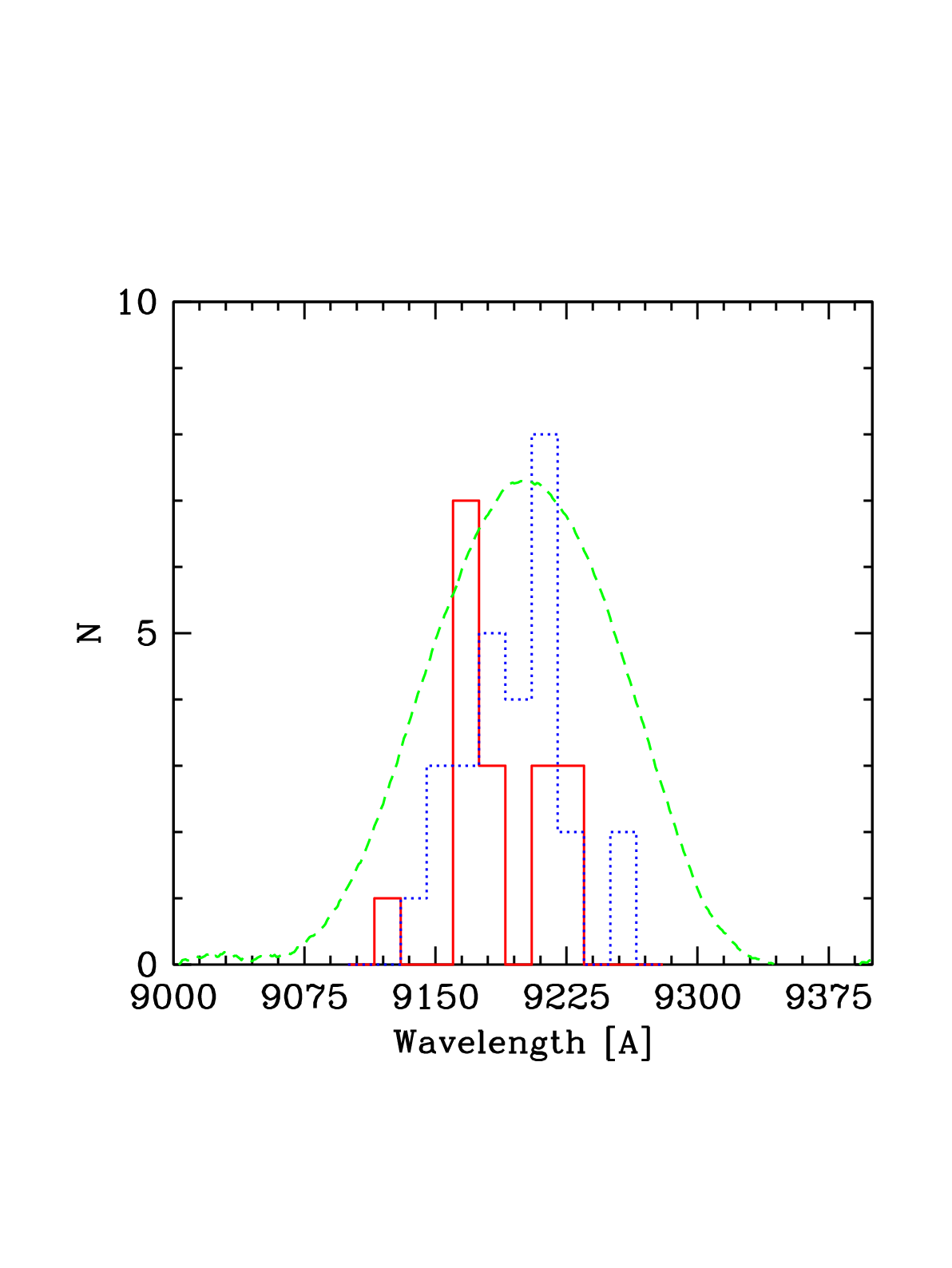}
\epsscale{1.0}
\caption{
Observed wavelength distribution of the Ly$\alpha$ line peak for our $17$ z6p5LAE sample (solid histogram), and our foreground sample (dotted histogram).
The transmission profile of $NB921$ is overplotted (dashed line).
\label{fig_peak}}
\end{figure}

We have also eight single-line emitters in which we detected neither an asymmetric line having as large $S_w$ as a LAE, nor doublet features as in the [O {\sc ii}] emitters.
Probably most of these are unresolved [O {\sc ii}] doublet lines, based on their small $S_w$.
In fact, all of these single lines were observed by Subaru FOCAS, whose data lacked the resolving power to separate the [O {\sc ii}] doublet ($\Delta\lambda=6.64$ \AA) at $9160$\AA.
Our $58$ photometric LAE candidates down to $NB921=26.0$ ($5$ $\sigma$) were selected in T05 based on the criteria that $z'-NB921>1$, $z'-NB921>3$ $\sigma$ and $i'-z'>1.3$ at $i'\leq27.87$ ($2$ $\sigma$) and simply $z'-NB921>1$ at $i'>27.87$ ($2$ $\sigma$).
We also adopted another criterion for LAE candidates with no detections ($<3$ $\sigma$) in deep $B$-, $V$-, and $R$-band images.
The $53$ objects for which we obtained spectra are composed of LAE candidates and $NB921$-strong emitters.
Twenty-two of these $53$ objects meet the criteria for our photometric sample: $16$ are LAEs, one is an $[$O {\sc iii}$]$ emitter, and five are single-line emitters.
We have another spectroscopically identified LAE that was not listed in the LAE candidate list by T05. 
The object is listed as No. $59$ in Table~\ref{tab_laespec} and is found to have a very close neighbor in the $i'$-band image (see Fig. \ref{fig_thum}), which prevented accurate aperture photometry.
Sixteen LAEs meet our selection criteria out of our spectroscopic LAE sample of $17$, indicating $16/17=94\%$ sample completeness, provided that all eight single-line emitters are foreground objects.
Otherwise, the sample completeness is $(16+5)/(17+8)=84\%$ at most, if all of the single emitters are z6p5LAEs.
The simple average of these two extreme cases, $89\%$, is regarded as the sample completeness.
On the other hand, the sample contamination rate is estimated as follows.
One $[$O {\sc iii}$]$ emitter satisfies our LAE criteria,\footnote{No. $29$ in T05.
This object is located near a bright star so that accurate photometry is prevented.} suggesting a $1/22=4.5\%$ contamination rate. 
The contamination rate could be $(1+5)/22=27\%$ at most if all of these five single emitters were foreground objects.
The contamination rate is estimated to be $16\%$ by taking the average of these two cases.
Therefore, our sample reliability factor, determined as the ratio of the number of true LAEs to the number of objects that meet our selection criteria, is evaluated to be $(1-0.16)/0.89=0.94$.
One object found to be an apparent $[$O {\sc iii}$]$ emitter by spectroscopy was removed from the photometric sample, whereas one LAE that happened to be found by spectroscopy but was not listed in the original candidate sample of T05 was included in the photometric sample.
In the following analysis, we used this photometric sample.
There were six objects for which we did not obtain any signals in spectroscopy.
Five of them have $NB921 > 25.5$, which is close to the current spectroscopic limit.

Figure~\ref{fig_peak} shows the peak wavelength distribution of $17$ confirmed LAEs, as well as $28$ foreground emitters within the $NB921$ bandpass.
The distribution of LAEs shows an apparent systematic deviation to the blue side of the NB transmission curve, in contrast to the symmetric distribution of foreground emitters.
This is because of the fact that the LAE profile, having a broad red wing and sharp blue cutoff, as well as a Lyman continuum break, makes a larger NB excess when it lies at the shorter side of the transmission curve.
This was also the case for LAE surveys at $z=5.7$ with $NB816$ \citep{shi06, hu04}.

\section{Ly$\alpha$ Luminosity Function}


We estimated the LF of z6p5LAEs based on both our spectroscopic sample of $17$ and our photometric sample of $58$.
The LF can be simply derived from the number density of confirmed LAEs multiplied by the spectroscopic confirmation fraction \citep[T05]{mal04}; however, the uncertainty in this fraction inevitably depends on the magnitude because spectroscopic confirmation is more difficult at fainter magnitudes.
Thus, we estimated the range of acceptable z6p5LAE LFs as specified by the upper and lower limits.

\epsscale{1.25}
\begin{figure}
\hspace*{-0.5cm}
\plotone{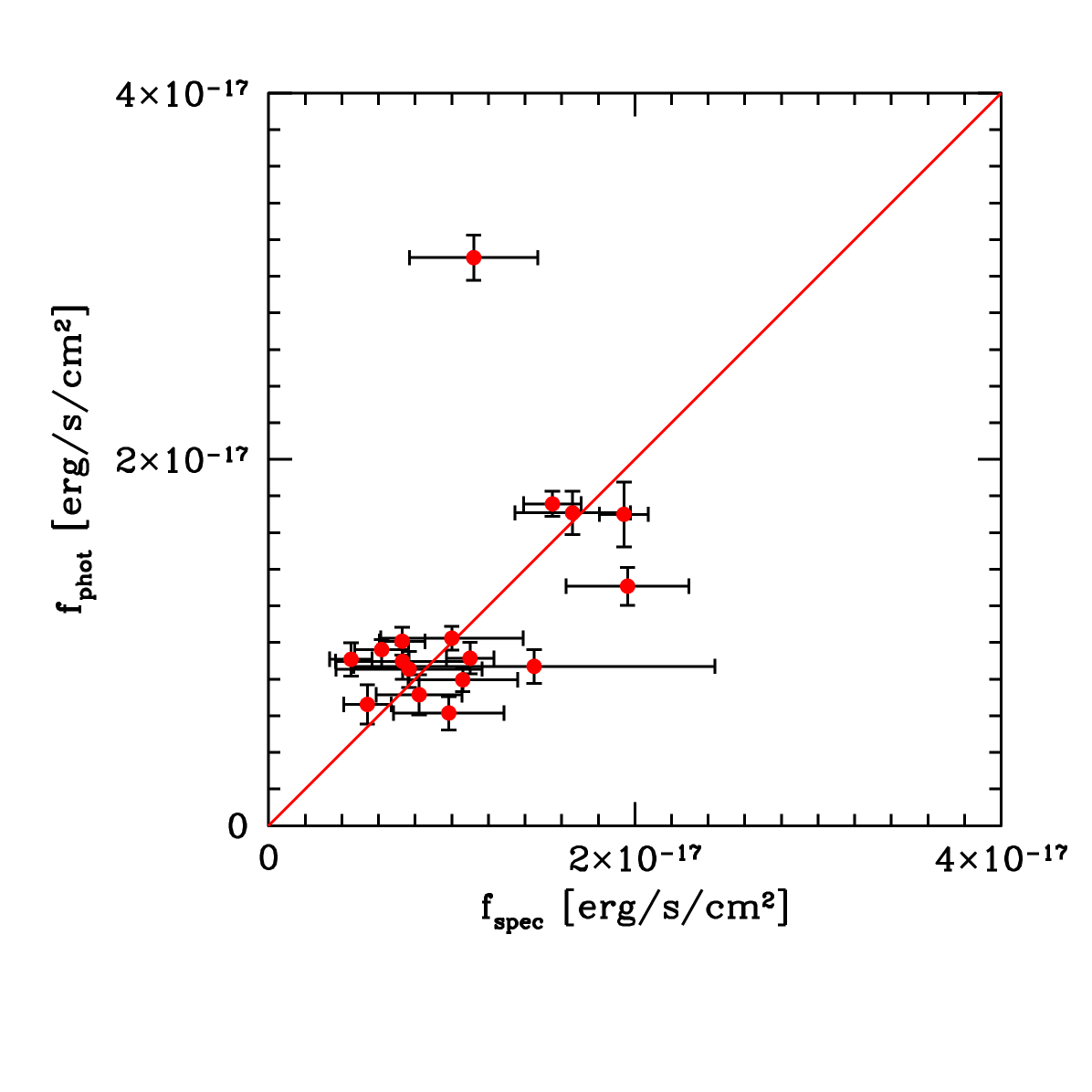}
\epsscale{1.0}
\caption{
Comparison of the Ly$\alpha$ fluxes measured by spectra, $f_{spec}$, with those inferred from photometry, $f_{phot}$, for our spectroscopic z6p5LAE sample.
The solid line represents a one-to-one correspondence between f$_{spec}$ and f$_{phot}$.
The errors were estimated based on the sky rms fluctuation on each spectra for f$_{spec}$, and errors on magnitudes for f$_{phot}$.
One object far out of agreement is found to have a spectrum affected by bad columns.
\label{fig_flaecomp}}
\end{figure}

\epsscale{1.4}
\begin{figure}
\hspace*{-0.5cm}
\plotone{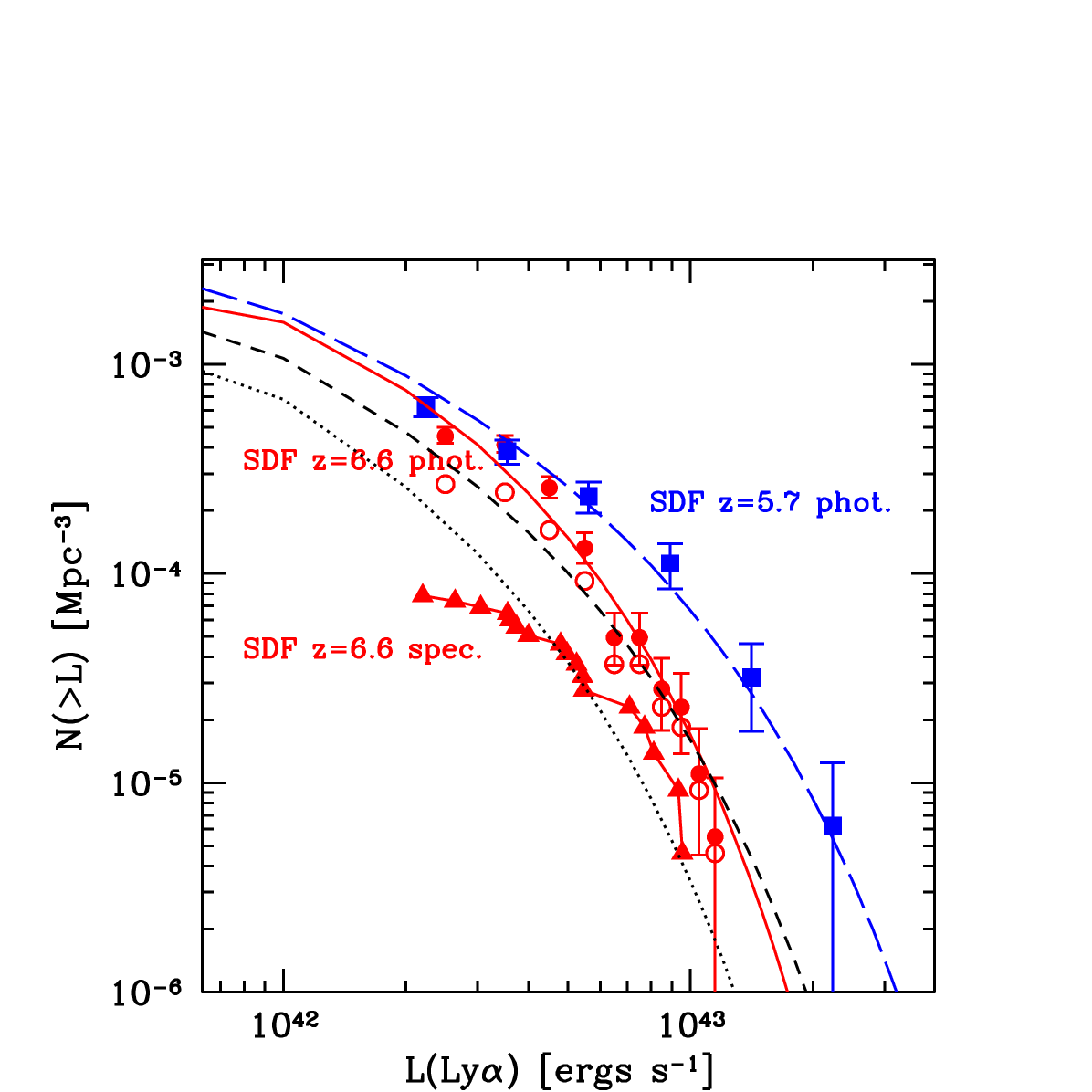}
\epsscale{1.0}
\caption{
Cumulative Ly$\alpha$ LF of our z6p5LAE sample.
The open circles denote the raw counts of our spectroscopic sample $+$ additional photometric sample, and the filled circles are corrected for detection completeness (upper limit).
The triangles denote the raw counts of the pure spectroscopic sample (lower limit).
Error bars for the filled circles are Poissonian.
The squares and long-dashed line indicate the LF of LAEs at $z=5.7$ evaluated from the SDF \citep{shi06}. 
The short-dashed and dotted lines show the Schechter LFs, in which the Ly$\alpha$ luminosities are reduced by a factor of $0.6$ ($L^*\times0.6$) and $0.4$ from the $z=5.7$ LF, respectively.
[{\it See the electronic edition of the Journal for a color version of this figure.}]
\label{fig_laelf}}
\end{figure}

The lower limit to the z6p5LAE LF is based on our spectroscopic z6p5LAE sample of $17$, whose Ly$\alpha$ emission has been securely detected, although this sample is incomplete.
The upper limit was estimated from the combined spectroscopic and photometric samples.
The Ly$\alpha$ and rest-UV continuum (at $9500$ \AA~) fluxes of our photometric sample were inferred from the $NB921$ and $z'$-band photometry using equations (7) and (6) of T05, respectively.
Note that we derived the apparent Ly$\alpha$ luminosity uncorrected for either the dust extinction or the self-absorption evident on the blue-side cutoff of the emission line.
The comparison of Ly$\alpha$ fluxes measured spectroscopically with those inferred from photometry for our spectroscopic z6p5LAE sample is shown in Figure~\ref{fig_flaecomp}.
The correspondence is good except in a few cases.
One object far out of agreement is found to have a spectrum affected by bad columns.
We therefore used the photometric inferred Ly$\alpha$ fluxes for the remaining $41(=58-17)$ objects in the photometric sample.

We have to correct for detection completeness, which could crucially affect the result when calculating the LF based on a deep photometric catalog.
The detection completeness as a function of apparent $NB921$ magnitude was estimated in almost the same way as in \citet{kas04}, that is, by counting detected artificial objects distributed on the real $NB921$ image.
We assumed Gaussian profiles of FWHM$=1\arcsec.13$ for these artificial objects, which is the nominal size of our z6p5LAE sample objects (T05).
The detection completeness was thus evaluated as $>75\%$ for $NB921<25.0$ and $45\%$ at the limiting magnitude of $NB921=26.0$.
In the upper limit estimate, we corrected for this detection completeness by number weighting according to the $NB921$ magnitude.
Note that the upper limit can be regarded as our current best estimate for the z6p5LAE LF, because it is properly corrected for the detection incompleteness.

Figure~\ref{fig_laelf} shows the cumulative Ly$\alpha$ LF of our z6p5LAE sample.
The open circles denote the raw counts of the spectroscopic $+$ additional photometric sample, and the filled circles are those with corrected detection completeness.
The triangles denote the raw counts of the pure spectroscopic sample.
Therefore, the triangles and circles are the lower and upper limits, respectively, of our estimates of the z6p5LAE LF.
The error bars on the filled circles just denote the Poisson errors, although there may be other plausible error sources, such as an ambiguity in inferring the Ly$\alpha$ luminosity from photometric data.
Taking into account the corrections with respect to the sample reliability factor ($94\%$) evaluated in the previous section, the LF has a margin to go upward by a factor of $1.06$, although this uncertainty is smaller than the Poisson errors.
We fitted a Schechter function, $\phi(L)dL=\phi^*(L/L^*)^\alpha$ exp$(-L/L^*)dL/L$, to our z6p5LAE LF.
The $\chi^2$ was minimized with a single grid search in the two parameters, $L^*$ and $\phi^*$, for fixed slopes of $\alpha=-2.0$, $-1.5$, and $-1.0$.
The derived best-fit parameters are listed in Table~\ref{tab_sch}.

The squares and dashed line in Figure~\ref{fig_laelf} show the LF of LAEs at $z=5.7$ evaluated in the SDF \citep{shi06}. 
This is the most reliable estimate so far because of the deeper photometric sample and larger number of spectroscopic confirmations than that of \citet{hu04}.
In addition, it is worth noting that this LAE sample at $z=5.7$ was selected from the same field and the same photometry as this study, having similar survey volume and similar selection criteria.
For example, we have carefully determined the NB-excess criteria so as to have almost the same equivalent width (EW$>20$ \AA~in rest frame) thresholds for both of these LAE samples. 

According to previous studies, the LF of LAEs between $z=3$ and $6$ is almost unchanged \citep{tra04, bre05}.
The LAE population appears to have a quite similar number density at all epochs.
A comparison of the LAE LF from previous studies at $z\leq5.7$ is shown in a companion paper by \citet{shi06}, which confirmed that the LF of LAEs at $z=5.7$ is almost identical to those in lower $z$.
However, our z6p5LAE LF shows an apparent deficit at the bright end in both upper and lower limit estimates.
The $L^*$ difference between $z=6.5$ (the upper limit) and $z=5.7$ is a factor of $\sim2$, corresponding to $\sim0.75$ mag, assuming a fixed $\alpha=-1.5$.
We have already spectroscopically identified almost all of our bright LAE candidates (four confirmed out of six candidates at $NB921\leq25.0$), and so this resulting deficit at the bright end would not change significantly, even if we obtain more follow-up spectroscopy in the future.
Moreover, uncertainties in the detection completeness correction and the selection effects according to the equivalent width of the Ly$\alpha$ line are expected to be small at the bright end.
As seen in Figure~\ref{fig_flaecomp}, there are possible but unsystematic errors in inferring the Ly$\alpha$ luminosity from photometric data, which could affect the resulting LF to some degree.
When using only the photometric inferred Ly$\alpha$ luminosities for our entire $58$ object sample, the best-fit Schechter parameters only change by at most $\Delta {\rm log}(L^*)=0.04$ and $\Delta {\rm log}(\phi^*)=0.1$ for $-2.0<\alpha<-1.0$, which is negligible.
Moreover, we carried out a Monte Carlo simulation to investigate any possible distortion that the discrepancy between spectroscopically measured and photometrically inferred Ly$\alpha$ luminosities could cause in the resulting LF. 
We re-calculated the LF many times after assigning a random error perturbed within the same scatter as in Figure~\ref{fig_flaecomp} to each Ly$\alpha$ luminosity.
With this simulation, the best-fit Schechter parameters vary with rms fluctuations of $\sigma( {\rm log}(L^*))=0.019$ and $\sigma( {\rm log}(\phi^*))=0.032$ for fixed $\alpha=-1.5$, suggesting again that the errors in inferring the Ly$\alpha$ luminosity are expected to be small.

\begin{deluxetable}{lllll}
\tabletypesize{\footnotesize}
\tablecaption{Best-fit Schechter Parameters for LAE LF at $z=6.5$ and $5.7$ of the SDF\label{tab_sch}}
\tablewidth{0pt}
\tablehead{
\colhead{Sample} & \colhead{Limit} & \colhead{$\alpha$} & \colhead{$L^*$}   & \colhead{$\phi^*$} \\
       & &         (fix)      & log(/$h_{70}^{-2}$ ergs s$^{-1}$) & log(/$h_{70}^{3}$ Mpc$^{-3}$)
}
\startdata
z = 6.5 & Upper & -2.0 & $42.74^{+0.14}_{-0.14}$ & $-3.14^{+0.30}_{-0.34}$ \\
      &      & -1.5 & $42.60^{+0.12}_{-0.10}$ & $-2.88^{+0.24}_{-0.26}$ \\
      &	    & -1.0 & $42.48^{+0.10}_{-0.08}$ & $-2.74^{+0.18}_{-0.22}$ \\
      & Lower & -2.0 & $43.30^{+0.23}_{-0.61}$ & $-4.80^{+1.02}_{-0.20}$ \\
      &      & -1.5 & $42.95^{+0.78}_{-0.42}$ & $-4.17^{+0.70}_{-0.83}$ \\
      &	    & -1.0 & $42.75^{+0.48}_{-0.32}$ & $-3.88^{+0.51}_{-0.57}$ \\
\tableline
z = 5.7 &     & -2.0 & $43.30^{+0.22}_{-0.18}$ & $-3.96^{+0.28}_{-0.30}$ \\
      &     & -1.5 & $43.04^{+0.12}_{-0.14}$ & $-3.44^{+0.20}_{-0.16}$ \\
      &	    & -1.0 & $42.84^{+0.10}_{-0.10}$ & $-3.14^{+0.14}_{-0.12}$ 
\enddata

\end{deluxetable}

At lower $z$ ($z=3.0-5.7$), several LAEs with large $L$(Ly$\alpha$)$>2\times10^{43}$ $h_{70}^{-2}$ ergs s$^{-1}$ have been actually identified by spectroscopy with much higher number density \citep{shi06, hu04, mai03, kud00, cow98}, whereas our z6p5LAE sample includes no spectroscopically confirmed objects of such large Ly$\alpha$ luminosity.
Three other LAEs at $z\sim6.5$ have been spectroscopically identified so far in independent surveys \citep{rho04, kur04, ste05} without taking advantage of the amplification by a foreground gravitational lens.
All of these LAEs also have similar Ly$\alpha$ luminosities of $L$(Ly$\alpha$)$=1.04-1.1\times10^{43}$ $h_{70}^{-2}$ ergs s$^{-1}$, consistent with our brightest LAEs.

On the other hand, at fainter luminosities of $L$(Ly$\alpha$)$<5\times10^{42}$ $h_{70}^{-2}$ ergs s$^{-1}$, our upper limit estimate of the z6p5LAE LF almost reaches the same amplitudes as that at $z=5.7$, although the amplitude difference between our upper and lower limits is too large to constrain its faint end.
Our faint spectroscopic sample is still too small to establish whether there is a significant difference between the faint end of the LF at $z=6.5$ and $5.7$.
In this study, we cannot determine the faint-end slope of the LAE LF at $z=6.5$, and consequently it is difficult to constrain the true contribution of the LAE population to the entire photon budget required for full reionization.

\epsscale{1.9}
\begin{figure}
\hspace*{-2.0cm}
\plotone{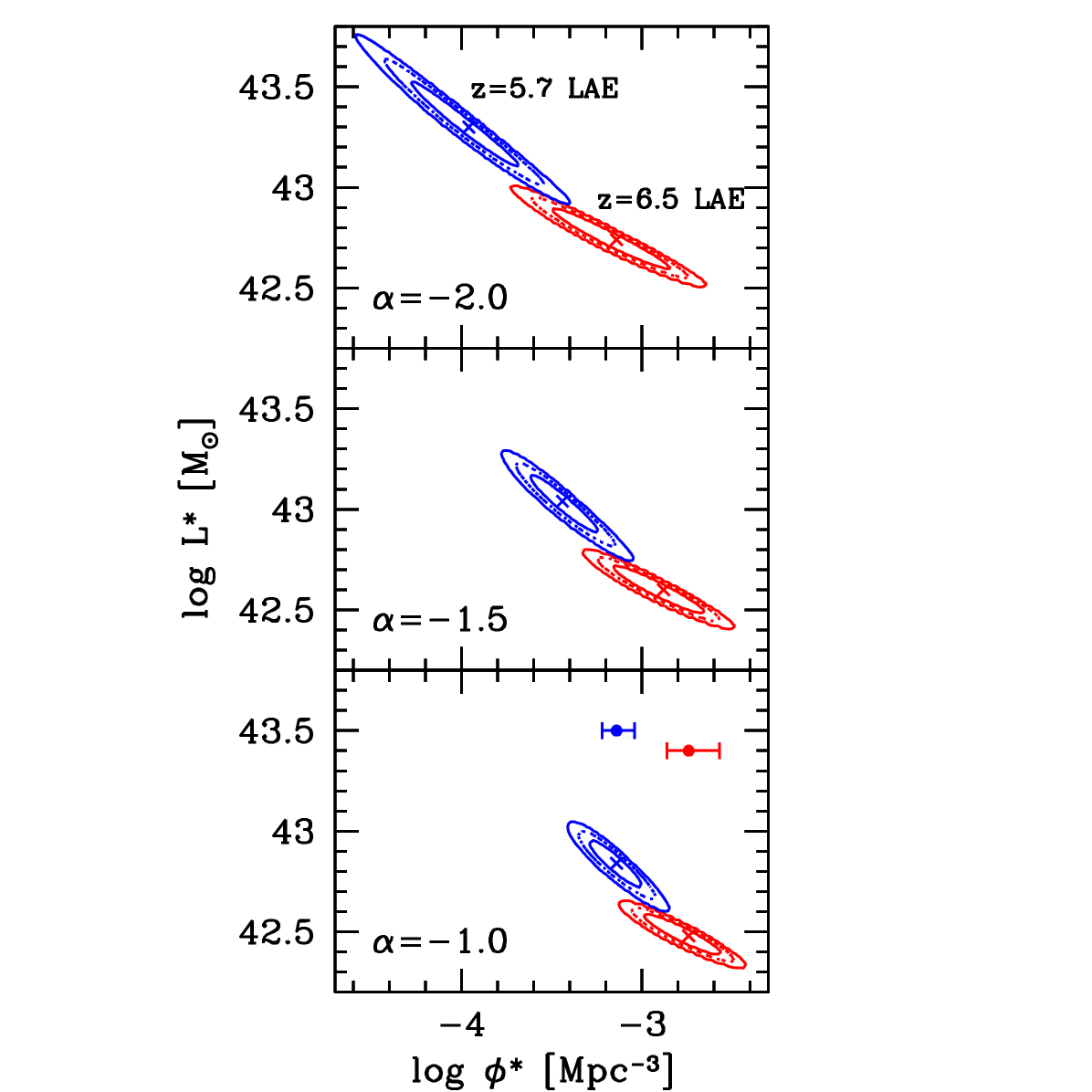}
\epsscale{1.0}
\caption{
Error ellipses of the best-fit Schechter parameters $\phi^*$ and $L^*$ for each fixed $\alpha$-value.
The lower set of ellipses in each panel are for the z6p5LAE upper limit sample, while the upper set of ellipses in each panel are for the z5p7LAE sample of \citet{shi06}.
The inner and outer solid ellipses are the $1$ $\sigma$ and $3$ $\sigma$ confidence levels, respectively.
The dotted ellipses show the $2$ $\sigma$ confidence levels.
The error bars in the lower panel ({\it left error bar}, z5p7LAE sample; {\it right error bar}, z6p5LAE sample) correspond to the uncertainty in $\phi^*$ due to cosmic variance \citep{som04}.
[{\it See the electronic edition of the Journal for a color version of this figure.}]
\label{fig_eelip}}
\end{figure}

\citet{mal04} and \citet{ste05} found no significant evolution of the LAE LF from $z=6.5$ to $5.7$.
However, their estimates were based on small samples combined from various independent data sets with different selection criteria. 
Our z6p5LAE sample was selected with the same criteria from a large homogeneous sample in a general field without resorting to amplification of gravitational lenses.
The uncertainty of our detection completeness estimate is small, at least at the bright end.
Moreover, our survey comoving volume is comparably as large as others at $z=5.7$ \citep{shi06, hu04}.
As seen in Figure~\ref{fig_peak}, our effective survey depth is smaller than that estimated from the FWHM of the NB filter; however, it is more or less the same as for the $z=5.7$ LAE sample, and its correction would not affect the result \citep{shi06}.
Therefore, our LF estimate of z6p5LAEs is highly reliable, although we cannot completely rule out the possibility that the deficiency at the bright end is caused by cosmic variance.

To illustrate the significance of the LF difference between $z=6.5$ and $5.7$, we plot the error contours for our Schechter-parameter fits in Figure~\ref{fig_eelip}.
In this case, we compare only the upper limit LF estimate, which is the current best estimate for our z6p5LAE sample and is most appropriate for establishing the significance of the LF difference compared to $z=5.7$.
The confidence levels of the fitting were computed based on Poissonian error statistics.
The best-fit parameters of the $z=5.7$ LF are slightly different from those presented in \citet{shi06}, in which Schechter parameters were determined so as to be consistent with the $NB816$ number count.
In contrast, here we determined these parameters simply by fitting a Schechter function to the data points.
Figure~\ref{fig_eelip} reveals that the ($L^*$, $\phi^*$) error ellipses for $z=5.7$ and $6.5$ do not overlap each other for any $\alpha$; that is, the difference in LF between $z=5.7$ and $6.5$ is significant at more than the $3$ $\sigma$ level.
The difference in $L^*$ is more significant than that in the $\phi^*$.
Based on \citet{som04}, we evaluated the cosmic variance of our z6p5LAE sample. 
We assumed a one-to-one correspondence between LAEs and dark haloes, and used their predictions at $z=6$.
With the comoving survey volume of $2.17\times10^5$ $h_{70}^{-3}$ Mpc$^3$ and the number density of $2.67\times10^{-4}$ $h_{70}^3$ Mpc$^{-3}$ ($7.83\times10^{-5}$) for the upper (lower) limit estimate, we obtained a cosmic variance of $\sim32\%$ ($\sim37\%$).
We also estimated a variance of $\sim 20\%$ for the $z=5.7$ LAE sample.
The possible field-to-field variance in the LF at $z=5.7$ can be seen in Figure 11 of \citet{shi06}.
As shown by the error bars in Figure~\ref{fig_eelip}, the $3$ $\sigma$ error circles for the two epochs overlap each other when taking into account the cosmic variance; however, our upper limit estimate still differs from the $z=5.7$ result at the $2$ $\sigma$ level.



\section{Rest-UV Continuum Luminosity Function}

In the previous section, the flux of the rest-UV continuum (at $9500$ \AA) was simultaneously derived from the $NB921$ and $z'$-band photometry.
We derived the rest-UV ($1255$ \AA~at $z=6.57$) continuum LF based on our photometric sample of $58$ objects.
The correction for detection incompleteness in $NB921$ was taken into account when calculating the LF, although the correction should actually be based on the completeness measured in $z'$ band, which corresponds to the rest-UV flux.
However, as our Ly$\alpha$-selected sample is basically constructed from an $NB921$ magnitude-limited sample, it is inevitable that the derived rest-UV continuum LF may be affected, especially at the faint end of the LF, by the difference in completeness of the $NB921$ and $z'$ bands.
This is also the case for other LAE studies \citep{shi06,hu04}.

\begin{figure}
\hspace*{-0.5cm}
\plotone{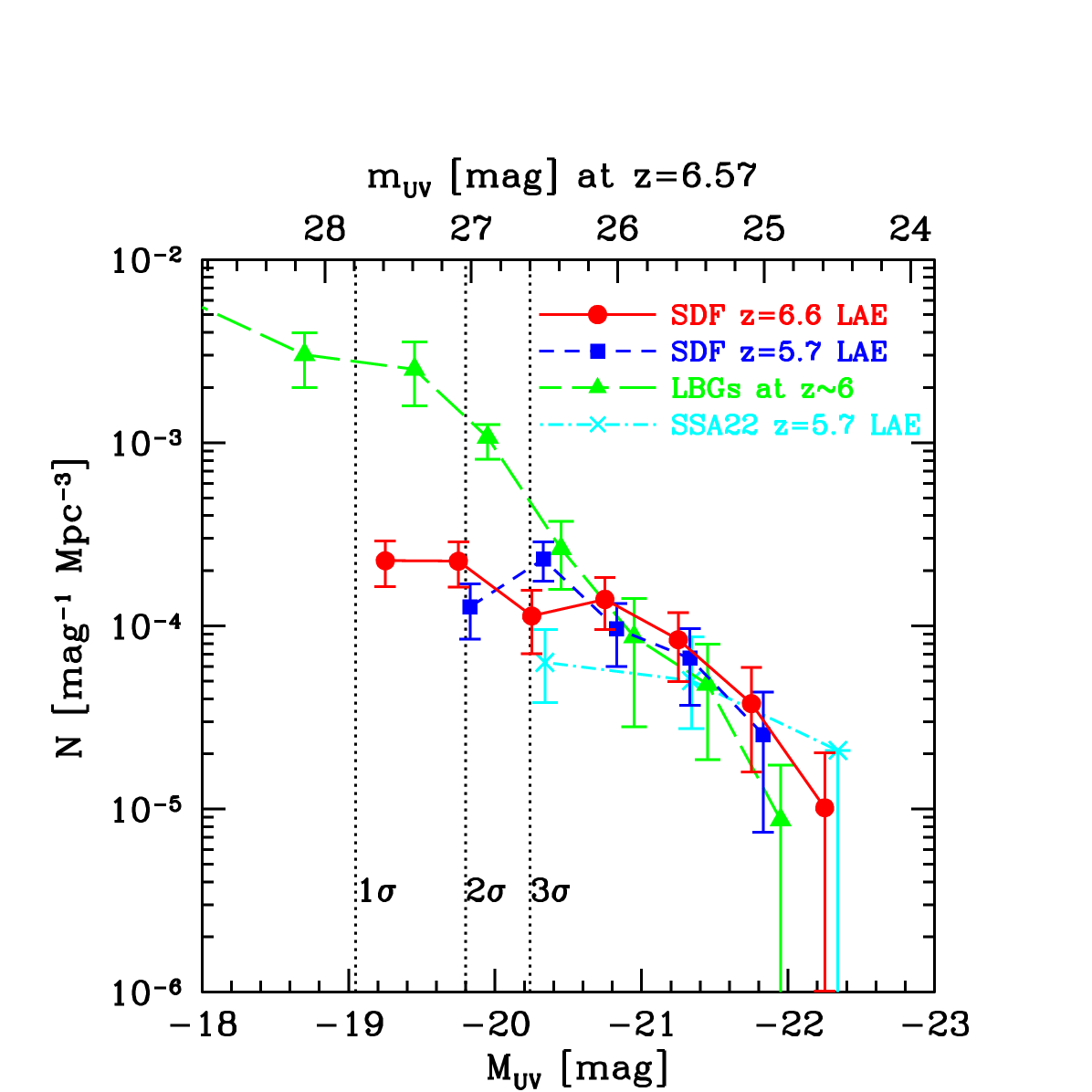}
\epsscale{1.0}
\caption{
Rest-UV ($1255$ \AA~at $z=6.57$) continuum LF of our z6p5LAE sample (circles and solid line).
As a comparison, the squares ({\it and short-dashed line}) are the rest-UV LF of the LAE sample at $z=5.7$ evaluated in the SDF \citep{shi06}, the triangles ({\it and long-dashed line}) are the $i$-dropout objects at $z\sim6$ of \citet{bou05}, and the crosses ({\it and dot-dashed line}) are the LAE sample at $z=5.7$ of \citet{hu04}.
The vertical lines indicate the limiting magnitudes in $z'$ band at $M_{UV}=-19.05$, $-19.80$ and $-20.24$ for $1$ $\sigma$, $2$ $\sigma$, and $3$ $\sigma$, respectively.
Error bars are Poissonian.
[{\it See the electronic edition of the Journal for a color version of this figure.}]
\label{fig_uvlf}}
\end{figure}

Figure~\ref{fig_uvlf} shows the rest-UV continuum LF of our z6p5LAE sample compared with other studies at similar redshifts.
No correction has been applied for dust.
The vertical lines indicate the corresponding limiting magnitudes in $z'$ band.
Our LF measurements at magnitudes fainter than $M_{UV}=-20.24$ ($3$ $\sigma$) may be uncertain because the corresponding $z'$-band magnitudes are no longer reliable.
We overplot in Figure~\ref{fig_uvlf} other rest-UV continuum LF estimates of the SDF LAE sample at $z=5.7$ \citep{shi06}, the $i$-dropout objects at $z\sim6$ of \citet{bou05}, and the LAE sample at $z=5.7$ of \citet{hu04}.
We neglect here a slight difference in corresponding rest-frame wavelengths ($\sim1350$ \AA~at $z=5.7$ and $\sim1255$ \AA~at $z=6.5$), assuming a flat far-UV spectral energy distribution.
Our measurements agrees with these two studies very well at $M_{UV}<-20.5$.
The agreement of the rest-UV continuum LF at the bright end for $z=6.5$ and $5.7$ is in clear contrast to the difference seen in the Ly$\alpha$ LF.
It should be noted that the rest-UV continuum luminosity is not attenuated by the neutral IGM and is less affected by dust extinction than the Ly$\alpha$ luminosity.
As far as the rest-UV continuum LF is concerned, cosmic variance is not severe for these samples.
The flatter faint-end slope of the LFs of LAE samples at both $z=6.5$ and $5.7$ compared to that of $i$-dropouts at $z\sim6$ could be caused by the detection incompleteness of the LAE sample.

\epsscale{1.4}

\section{Clustering Properties}

\begin{figure}
\hspace*{-0.5cm}
\plotone{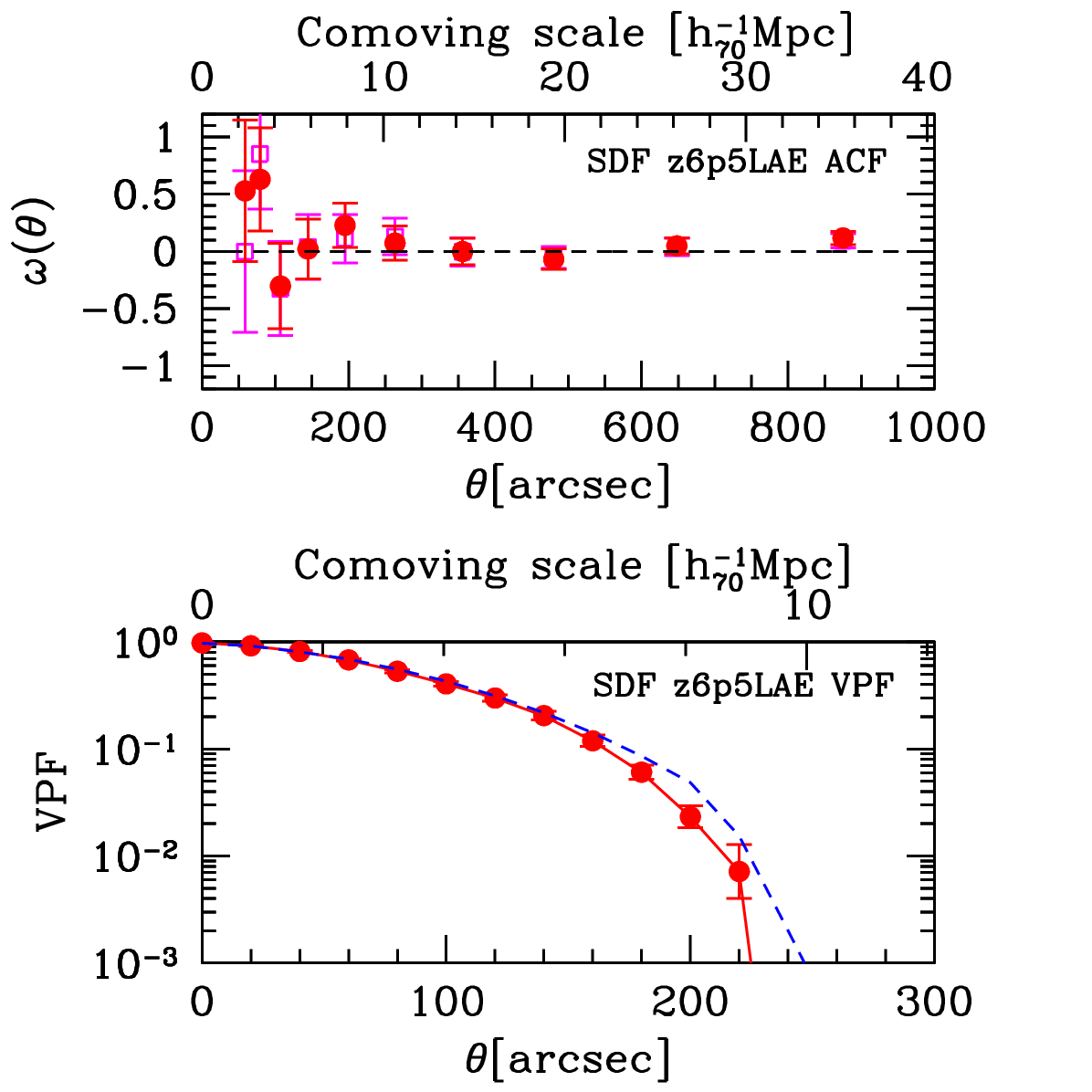}
\epsscale{1.0}
\caption{
{\it Top}: ACF of our z6p5LAE sample.
The circles denote the ACF for the entire z6p5LAE sample of $58$ objects, whereas squares denote that for the $53$ z6p5LAE sample in which five single-line emitters were removed.
Error bars show the $1$ $\sigma$ Poissonian errors.
{\it Bottom}: VPF of our z6p5LAE sample ({\it circles}).
The dashed line is the VPF of a random sample.
\label{fig_acfvpf}}
\end{figure}

Our sample of $58$ z6p5LAE candidates have been extracted from a very wide field of view ($34\arcmin\times27\arcmin$).
We tried to detect a clustering signal in the z6p5LAE sample using several methods.
We derived the angular two-point correlation function (ACF) $w(\theta)$ using the \citet{ls93} estimator.
One hundred thousand random points were created with exactly the same boundary conditions as the SDF galaxy catalog, avoiding the mask regions in which saturated stars dominate.
The top panel of Figure~\ref{fig_acfvpf} shows the ACF for z6p5LAE.
Circles denote the ACF for the $58$ objects of z6p5LAE, whereas squares denote the $53$ objects of the z6p5LAE sample, excluding five single-line emitters.
We did not correct for the integral constraint, which is negligible in the SDF \citep{kas06}.
We estimated only Poissonian errors on the ACF as $\sigma_w(\theta)=\{(1+w(\theta))/DD(\theta)\}^{0.5}$, where $DD(\theta)$ is the number of random-random pairs having angular separation $\theta$.
The result shows that the amplitude is almost zero for all scales, indicating that the sample has an almost homogeneous distribution.
However, our z6p5LAE sample is so small that the derived ACF has a large ambiguity.
Therefore, we also applied two other methods that are more robust for small number statistics to quantify the clustering strength.

\epsscale{1.4}

First, we applied the two-dimensional Kolmogorov-Smirnov test to our sample.
This test was generalized by \citet{pea83} to give the integral probability distribution in four quadrants around a certain point.
To see the difference from a homogeneous distribution, we generated random points as for our ACF estimate.
We found that our z6p5LAE sample was equivalent to a homogeneous distribution at the $83.3\%$ ($83.8\%$ after removing the five single-line emitters) confidence level.

Second, we estimated the void probability function (VPF).
The VPF is defined as the probability of having no galaxies in a randomly placed sphere of radius $R$, or in a circle of angular radius $\theta$ in the case of a two-dimensional sky distribution.
The VPF is known to be related to the hierarchy of $n$-point correlation functions \citep{whi79}.
We adopted the same technique as \citet{cro04} to correct for the irregular geometry of the survey region.
The result is shown as a solid line in the bottom panel of Figure~\ref{fig_acfvpf}, compared with that of a random sample indicated by the dashed line.
No excess in the VPF was found for our z6p5LAE sample relative to the random sample.

These three estimates indicate that the spatial distribution of the z6p5LAE sample is homogeneous.
The results were identical for all three estimates, even if we divided our sample into brighter/fainter subsamples.
The homogeneous distribution of z6p5LAEs is in contrast to that at lower $z$, where LAE populations are often found to trace the large-scale structures even in blank fields \citep{ste00,shi03,pal04,ouc05}.

As in the previous section, we estimated the cosmic variance as $\sim32\%$ at most.
We should note that NB searches exploring only a small redshift coverage are also sensitive to the large-scale structure \citep{shi04}.
We cannot rule out the possibility that we happen to see a very homogeneous region of the $z=6.5$ universe.
A survey of a much larger volume is required for further discussion.





\section{Ly$\alpha$ Profile of the Composite Spectrum}

Although each individual spectrum of our $17$ z6p5LAEs has too low a signal-to-noise ratio (S/N) to infer either the internal dynamics of the LAE itself or the IGM characteristics (e.g., \citealp{hai02}), the composite spectrum could be useful to see the general spectroscopic properties of the z6p5LAE population.
We have $17$ z6p5LAE spectra of different spectroscopic resolutions.
First, we removed five spectra\footnote{The removed sample were object ID=4, 5, 6, 7, and 8 in T05} that have been taken with the poorest instrumental resolution.
Each spectrum was then smoothed with an adequate Gaussian kernel chosen to produce a common instrumental resolution of FWHM$=6.41$ \AA.
The instrumental resolution for each spectrum was practically measured from the FWHM of sky lines near the Ly$\alpha$ emission.
The redshift was measured based on the line peak wavelength, then shifted to the mean redshift $\langle z \rangle=6.556$, and rebinned to a common pixel scale.
In the process, we neglected possible systematic offsets of Ly$\alpha$ lines from the rest frame established by other lines, which are often found in LBG spectra \citep{sha03}.
The spectra were then combined by taking the average after scaling according to their peak line flux.

\begin{figure}
\hspace*{-0.3cm}
\plotone{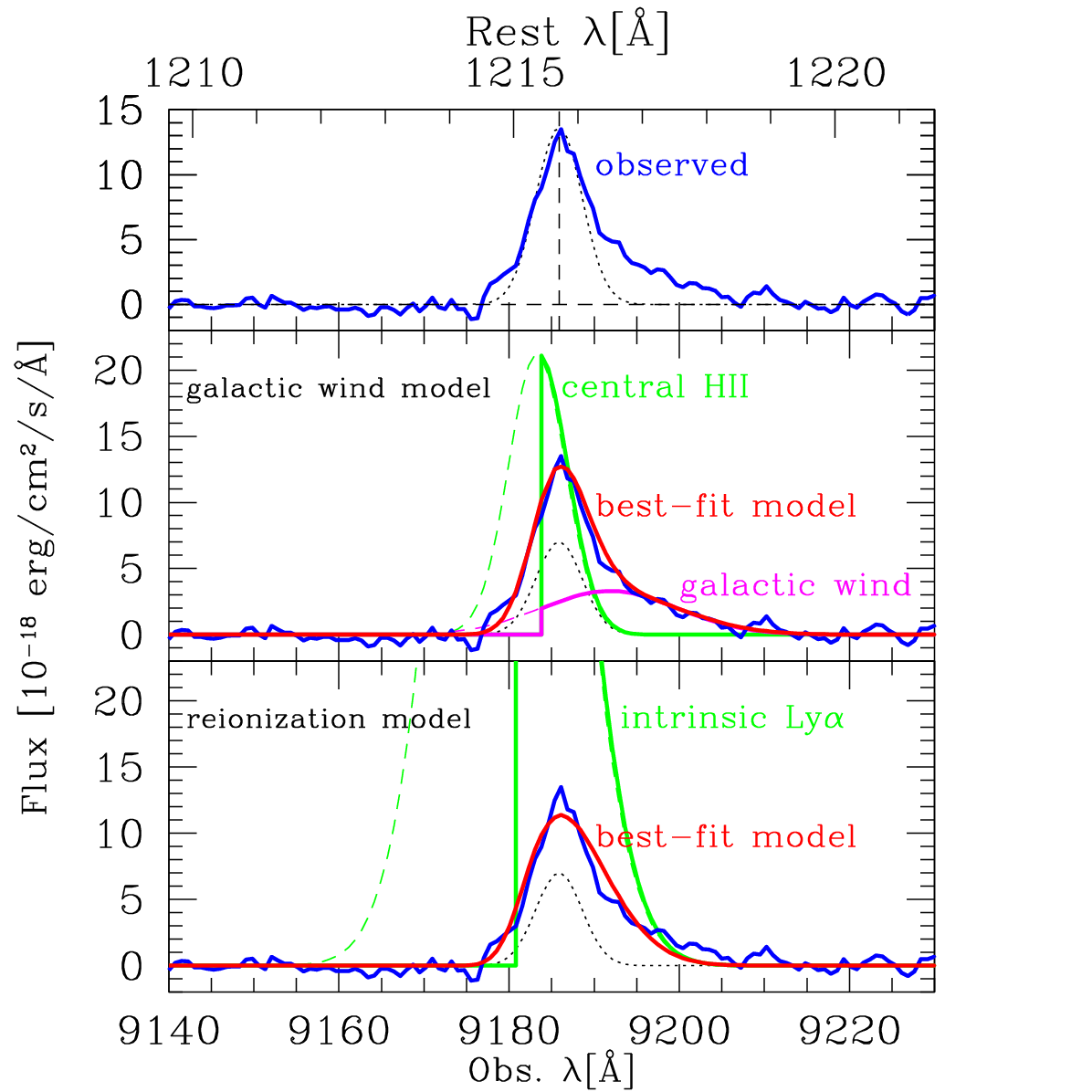}
\epsscale{1.0}
\caption{
The composite spectrum of our $12$ z6p5LAE sample.
The spectrum in the top panel shows the final composite spectrum.
The dotted line denotes a Gaussian profile corresponding to the instrumental resolution.
The middle panel shows the best-fit galactic wind model. 
The resulting profile is shown as a smooth solid line that is the combination of narrow and broad Gaussians convolved with the instrumental resolution.
The lower panel shows the reionization model. 
The resulting profile is shown by a smooth solid line that is the combination of the intrinsic Ly$\alpha$ profile and damping wing attenuation convolved with the instrumental resolution.
See Table~\ref{tab_prof_param} for the best-fit parameters for each model.
[{\it See the electronic edition of the Journal for a color version of this figure.}]
\label{fig_prof}}
\end{figure}

The top panel of Figure~\ref{fig_prof} shows the final composite spectrum from our $12$ z6p5LAE sample.
The composite spectrum reveals an apparently asymmetric profile with an extended red wing, which is shown clearly by a comparison with a Gaussian profile ({\it dotted line}) corresponding to the final instrumental resolution.
The skewness and weighted skewness of the composite spectrum were $S=0.542\pm0.007$, and $S_w=11.466\pm0.156$, respectively.
The blue side of the line profile is almost adequately explained by the instrumental resolution blur, as was also concluded by \citet{hu04} and \citet{wes05}.
Assuming that an intrinsic Ly$\alpha$ profile is a simple Gaussian at almost the same peak position as observed, and completely truncated at the blue side of the line, the resulting profile convolved with the instrumental resolution did not coincide with the observed profile, instead producing a large red-wing anomaly.
This disagreement is inconsistent with the results for the $z=5.7$ case by \citet{hu04}.

\begin{deluxetable*}{llcccl}
\tabletypesize{\footnotesize}
\tablecaption{Model Parameters for Composite Ly$\alpha$ Profile of LF of z6p5LAE\label{tab_prof_param}}
\tablewidth{0pt}
\tablehead{
\colhead{Model} & \colhead{Component} & \colhead{$\lambda_c$} & \colhead{Amplitude} & \colhead{$\sigma$} & \colhead{Offset} \\
                & &        (\AA)         & (ergs s$^{-1}$ cm$^{-2}$ \AA$^{-1}$) & (\AA) (km s$^{-1}$) & (\AA) (km s$^{-1}$) 
}
\startdata
Galactic wind model & Central H$_{\rm II}$   & 9183.3 & 2.11E-17 & $3.32$ $108.5$ & $-2.34$ $-76.5$ \\
                    & Galactic wind & 9191.6 & 3.31E-18 & $7.93$ $259.2$ & $+5.97$ $+195.1$ \\
\tableline
Reionization model &               & 9179.9 & 9.00E-17 & $6.50$ $212.4$ & $-5.70$ $-186.3$
\enddata

\end{deluxetable*}

\epsscale{1.4}

To explain the observed profile, we considered two plausible models.
The first model is the ^^ ^^ galactic wind model" which was motivated by the similar analogy with \citet{daw02} and \citet{mas03}.
If galactic winds are present, the far side of the expanding shell back-scatters redshifting Ly$\alpha$ photons that would make another broadly extended Gaussian component in their line profile.
We simply assumed that the Ly$\alpha$ photons from the blue side of the object redshift are completely absorbed by neutral hydrogen at the near side in lines of sight.
This model is composed of two Gaussian profiles: one is a high-amplitude narrow Gaussian that originates from recombination Ly$\alpha$ photons in the central H {\sc ii} region, and the other is a low-amplitude broad Gaussian from Ly$\alpha$ photons back-scattered by a galactic wind.
The middle panel of Figure~\ref{fig_prof} shows the best-fit galactic wind model.
The resulting profile is shown by the solid line, which is the combination of narrow and broad Gaussians convolved with the instrumental resolution.
It perfectly explains the observed profile.
The best-fit parameters are listed in Table~\ref{tab_prof_param}.
The picture of large-scale outflowing of gas with velocities $\sim200$km s$^{-1}$ is in good agreement with those of nearby H {\sc ii} galaxies \citep{kun98} but is somewhat smaller than those of $z\sim3$ LBGs \citep{pet02} and $z\sim5$ LAEs \citep{daw02,wes05}.

The second model is the ^^ ^^ reionization model", in which the intrinsic Ly$\alpha$ line has a larger amplitude than that observed and its peak wavelength is much shorter than the observed peak position.
Such a picture is generally predicted for the Ly$\alpha$ profile in the reionization epoch \citep{hai02, san04}.
The Ly$\alpha$ photons would be absorbed by both the red damping wing of the Gunn-Peterson trough from outside the cosmological H {\sc ii} region and the residual neutral hydrogen inside the H {\sc ii} region.
The latter's true contribution is not yet known; thus, we simply assumed that the inside of the H {\sc ii} region is sufficiently optically thick to completely attenuate the blue side of the line. 
The damping wing scattering would be effective at wavelengths larger than that corresponding to the blue edge of the H {\sc ii} region; thus, the red side of the line would be attenuated by the damping.
We used the damping optical depth of \citet{loe05}.
We further assumed the radius of the H {\sc ii} region to be $0.45$ proper Mpc \citep{hai02, san04}.

The bottom panel of Figure~\ref{fig_prof} shows the cosmological H {\sc ii} region model fitting to the data, and the best-fit parameters are listed in Table~\ref{tab_prof_param}.
This model also explains the observed extended red wing fairly well.
The predicted intrinsic Ly$\alpha$ luminosity from this model is $4.6$ times that observed, which is roughly consistent with the case of \citet{hu02} and its model prediction by \citet{hai02}.
However, the transmission factor of $\sim20\%$ of the total line flux is smaller than the factor of $\sim50\%$ dimming suggested by the LF.
Here, we neglect the possible luminosity dependence of Ly$\alpha$ attenuation, which was suggested by the LF shown in \S~3.
The discrepancy in Ly$\alpha$ attenuation between that implied by the line-profile model fitting and that from the LF difference can be reduced by assuming a larger radius for the H {\sc ii} regions.
For example, we can obtain a $\sim40\%$ transmission factor if the radius of the H {\sc ii} region is as large as $\sim 0.90$ proper Mpc.
Although we assume an optically thick core for the H {\sc ii} region, any escape of Ly$\alpha$ flux at the blue side would broaden the line profile, making the fit worse. 
We conclude that a larger contribution of the residual neutral hydrogen inside the H {\sc ii} region compared to the damping wing is required, although our data still lack the spectral resolution to make quantitative predictions of the density profile inside the H {\sc ii} region.

A better fit is obtained for the galactic wind model; however, our composite spectrum still has too low S/N and spectral resolution to determine which model is more plausible.
It should also be noted that there is likely to be a scatter in FWHM among our LAE sample, and so it is unclear whether all of the LAEs have prominent red wings that appeared in the composite spectrum.
While in the model, the radiative transfer process of Ly$\alpha$ photons through the emitting galaxy and the IGM is too complicated to justify every parameter based only on the line-profile fitting.


\section{Implications for Reionization}

In this study, we found that the z6p5LAE LF has a clear deficit at its bright end compared with that at $z=5.7$.
The simplest interpretation is that the LAE population undergoes Ly$\alpha$ luminosity evolution from $z=6.5$ to $5.7$.
Strictly speaking, the LAE population has some evolution in its EW from $z=6.5$ to $5.7$, given that we observe no evolution in the rest-UV continuum LF.
The LAEs must be a very young population, having ongoing starbursts in so short period that even $100$ Myr is a long time, over which their Ly$\alpha$ luminosity can easily drop.
Nevertheless, the number density of LAEs does not change from $z\sim3$ to $\simeq5.7$.
Thus, it is more natural to assume that this lack of LF evolution should extend up to $z=6.5$, as opposed to the LAE population having strong evolution between $z=5.7$ and $6.5$.
The number density decline from $z=5.7$ to $6.5$ could imply a substantial transition in the cosmic ionization state between these epochs.
In this section, we offer a possible interpretation of these observational results in the context of the reionization of the universe, assuming that the nature of LAEs themselves has no drastic evolution between $z=5.7$ and $6.5$.

Assuming a fully ionized IGM at $z=5.7$, the comparison of the LFs at $z=5.7$ and $6.5$ puts constraints on the neutral fraction of IGM hydrogen $x_{\rm H I}^{\rm IGM}$ at $z=6.5$.
The short-dashed and dotted lines in Figure~\ref{fig_laelf} show the Schechter LF, in which the Ly$\alpha$ luminosities are reduced by a factor of $0.6$ ($L^*\times0.6$) and $0.4$ from the LF at $z=5.7$, which is still consistent with our upper and lower limit LF estimates, respectively.
According to the IGM dynamical model of \citet{san04}, a Ly$\alpha$ luminosity drop by $\Delta L^*\sim0.75$ from the fully ionized IGM corresponds to $x_{\rm H I}^{\rm IGM}=0.45$.
However, this predicted value is strongly model dependent, and even the model of \citet{san04} covers a wide range of acceptable models with different predictions of Ly$\alpha$ attenuation.
The predicted value of $x_{\rm H I}^{\rm IGM}$ can be much smaller than $0.45$ in some of these models.
Therefore, our LF estimate could allow a neutral fraction of the IGM at $z=6.5$ of $0 \lesssim x_{\rm H I}^{\rm IGM} \lesssim 0.45$.
This upper limit of $x_{\rm H I}^{\rm IGM}$ at $z\sim6.5$ is consistent with the recent results of \citet{mal06} and \citet{tot06}.

\citet{hai05} evaluated the LF evolution of LAEs, taking into account the luminosity dependence of the Ly$\alpha$ flux attenuation.
In the epoch of reionization, ionizing sources like LAEs would make surrounding cosmological H {\sc ii} regions \citep{mir00}.
Based on the CDM model, the galaxies embedded in massive dark halos would collapse first, so that the ionizing sources in this era would be preferentially located in the high-density regions.
The H {\sc ii} regions of the bright LAEs clustered in the overdense regions would overlap effectively and create a larger H {\sc ii} region with a high ionization fraction, which would significantly reduce the Ly$\alpha$ flux attenuation.
As a result, it is predicted that bright LAEs should be readily observed, whereas faint LAEs are more severely attenuated.
However, this luminosity dependence would just shift the Schechter LF downward by a certain factor according to $x_{\rm H I}^{\rm IGM}$ because the LF is steeper at the bright end.
Our observed spectroscopic LF (lower estimate) is nearly consistent with such a trend, but our upper limit estimate of the z6p5LAE LF, which has a steep decline only at its bright end, is not.
Our null result of finding any signals of bright LAE clustering is also inconsistent with this model.
As \citet{hai05} suggested, the predicted LF profile strongly depends on the model assumptions,
 such as the constant escape fraction with respect to Ly$\alpha$ luminosity.

Assuming a clumpy IGM and discrete ionizing sources, there are two conflicting model predictions about the spatial inhomogeneity of reionization propagation.
In a first phase of reionization, most of the sources are formed in high-density regions and ionize the dense gas around them.
If the local neutral IGM around these sources is dense enough to allow higher recombination rates than ionization rates, the reionization will not complete in these high-density regions.
Consequently, as in the model prediction by \citet{mir00}, H {\sc ii} regions would expand preferentially towards low gas density regions, and overdense regions would gradually be ionized after the epoch of overlap of cosmological H {\sc ii} regions (^^ ^^ outside-in" model).
This picture is supported by hydrodynamic \citep{gne00} and $N$-body \citep{cia03} simulations.
On the other hand, if the local IGM density is low enough to help an efficient escape of Ly$\alpha$ photons, it is expected that the dense IGM region, in which a large H {\sc ii} region can be formed by overlap of ionized bubbles around high-luminosity sources, would be ionized first.
It would then proceed to a void where it is dominated by only low-luminosity sources (^^ ^^ inside-out" model).
Such a picture is predicted by \citet{sok03} and \citet{fur06}.

The key diverging point between these two models is the IGM density distribution, which \citet{mir00} assume as an extrapolation from that at $z=3$; however, the adequacy of that assumption is still uncertain \citep{fur06}.
These two contradictory models obviously have different predictions about the spatial distribution of observable galaxies at that epoch: galaxies residing in the underdense regions are easily observed in the outside-in model, whereas galaxies residing in the overdense regions are easily observed in the inside-out model.
In turn, a more inhomogeneous galaxy distribution is predicted in the inside-out model.
Assuming that we are seeing the final stage of reionization through our observed z6p5LAE sample as deduced from the previous estimation of $0 \lesssim x_{\rm H I}^{\rm IGM} \lesssim 0.45$, the homogeneous distribution found in this work agrees better with the outside-in model.
Our assumed picture, in which the LAE overdense regions were still obscured by surrounding thick neutral IGM clouds at $z=6.5$, is also consistent with our LF deficit in bright LAEs.
Such an overdense region could be kept neutral until later epochs, trapping bright ionizing sources like luminous QSOs, in which the appearance of Gunn-Peterson troughs are detected at lower $z$: $z\sim5.2-5.8$ \citep{bec01, djo01}.
Inhomogeneous reionization is also suggested by the significant variation in the IGM transmission among different QSO lines of sight \citep{djo06}.
In summary, implications for the reionization process suggested by this study are a high clumping factor of the IGM, and inhomogeneous reionization.
Although we do not draw any further quantitative conclusions, we may be looking at only low-luminosity LAEs residing in low-density IGM regions at the end of the reionization epoch.

\epsscale{1.4}
\begin{figure}
\hspace*{-1.2cm}
\plotone{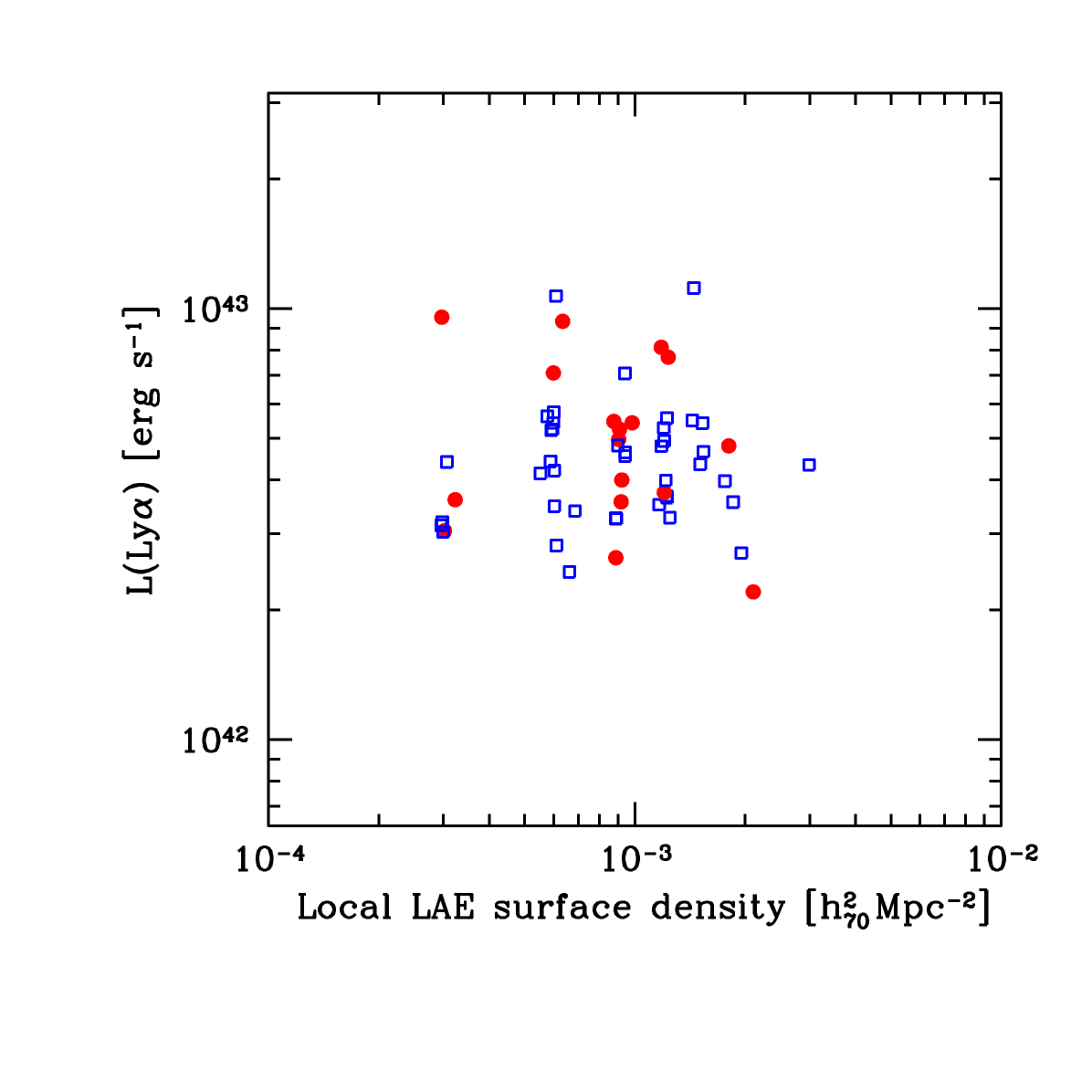}
\epsscale{1.0}
\caption{
Correlation between the Ly$\alpha$ luminosity and the local surface number density for our $58$ z6p5LAE sample.
The circles are from the spectroscopic sample, whereas the squares are from the photometric sample.
\label{fig_llae_dens}}
\end{figure}

We plot in Figure~\ref{fig_llae_dens} the relation between the local surface number density of LAEs and their Ly$\alpha$ luminosities, $L$(Ly$\alpha$).
The local surface number density is measured by the number of LAEs in a circle of $8h_{70}^{-1}$ Mpc comoving radius around each sample object.
There is no apparent correlation between the local density and $L$(Ly$\alpha$), again suggesting spatial homogeneity.
It should be noted that the local number density in Figure~\ref{fig_llae_dens} just accounts for our LAE sample, and at this point, there is no way of inferring the presence of other ionizing sources like $i'$-dropout galaxies without strong Ly$\alpha$ emission.
Nor did we find any clear relation between the local densities and the FWHM of Ly$\alpha$ lines.

\epsscale{1.4}
\begin{figure}
\hspace*{-1.4cm}
\plotone{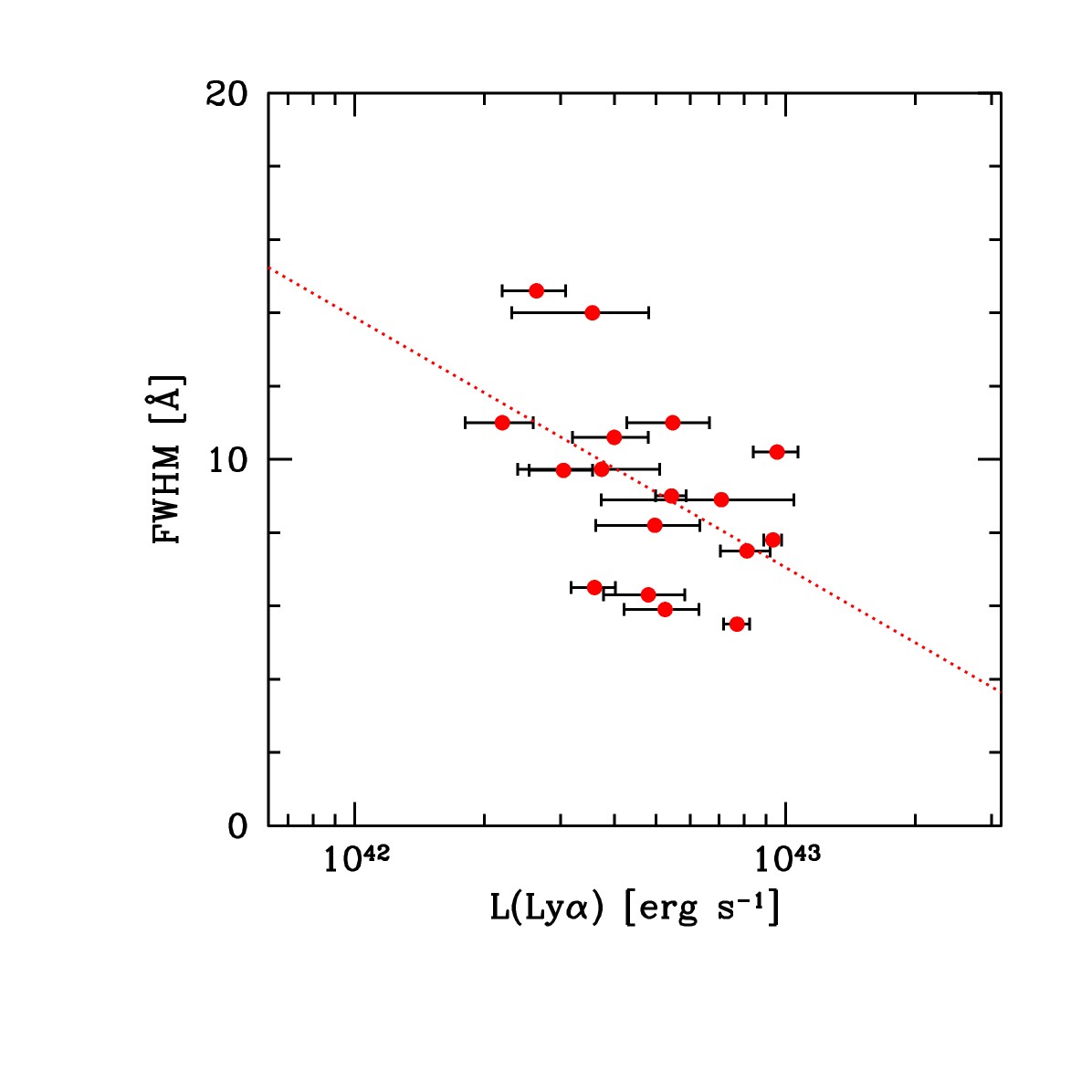}
\epsscale{1.0}
\caption{
Correlations between the Ly$\alpha$ luminosity and the FWHM of the Ly$\alpha$ emission line for our sample of $17$ spectroscopically confirmed z6p5LAEs.
The dotted line is the fiducial linear fit.
\label{fig_llae_fwhm}}
\end{figure}

Figure~\ref{fig_llae_fwhm} shows the FWHM of Ly$\alpha$ emission as a function of $L($Ly$\alpha)$ for our spectroscopic z6p5LAE sample.
There is a weak anti-correlation between FWHM and $L($Ly$\alpha)$.
\citet{hai05} predicted a similar correlation, which they proposed as an independent diagnostic of $x_{\rm H I}^{\rm IGM}$, aside from the LF.
The blue side of Ly$\alpha$ emission is dominantly attenuated by the residual H {\sc i} inside the cosmic H {\sc ii} regions.
A systematic high velocity, as in galactic winds, would reduce more effectively the amount of residual H {\sc i} for smaller H {\sc ii} regions surrounding low-luminosity sources; thus, the anti-correlation is expected.
This effect would depend more sensitively on the line width at the blue side, which is difficult to measure precisely.
The correlation is expected to be steeper for higher $x_{\rm H I}^{\rm IGM}$; however, the observed correlation in Figure~\ref{fig_llae_fwhm} shows too much scatter to determine $x_{\rm H I}^{\rm IGM}$.
The observed anti-correlation could be a sign of a high $x_{\rm H I}^{\rm IGM}$, indicating that cosmic reionization has not been completed at $z=6.5$.
A possible variation in systematic internal velocity among LAEs would also dilute the relation.
The intrinsic relation between FWHM and $L($Ly$\alpha)$ has not been clearly established.
\citet{mat05} found a clear positive correlation between the velocity dispersion and $L($Ly$\alpha)$ for their large Ly$\alpha$ blob sample at $z\sim3$ in the SSA22 proto-cluster region.
The intrinsic relation should be determined for the general LAE population at lower $z$, where it is irrelevant to the reionization.




The possible implications for reionization described above are still speculative because there are many unknown factors.
First, we did not take into account the inherent galactic evolution of LAEs. 
It is suggested by previous studies that the number density of LAEs between $z=3$ and $6$ is almost unchanged, which is in contrast to the LBG population, which undergoes significant LF evolution from $z=3$ to $6$ \citep{ouc04,bou05,yos06}.
However, less than $20$ LAEs have been confirmed spectroscopically, even at $z\sim3$; this is not enough to determine the LF accurately.
On the other hand, \citet{bou05} have found an evidence for evolution of the rest-frame continuum UV LF between $z\sim3$ and $6$.
Also, there is no consensus on the spatial distribution of LAE at low redshifts. 
They might have an intrinsically homogeneous distribution.
The inherent evolutions of Ly$\alpha$ flux, dust content, and neutral gas fraction inside LAEs themselves would also decrease the number density from $z=5.7$ to $6.5$.
The complex escape mechanism of ionizing radiation from galaxies is unclear and depends strongly on assumed parameters, such as the escape fraction of Ly$\alpha$ photons, amount of dust, galactic wind, and star formation activity.
Our observed z6p5LAE LF could be more consistent with the quiescent hierarchical model prediction by \citet{del05} than with their model including an ongoing starburst.
Finally, the IGM physical conditions during the reionization epoch are unknown, and no observational evidence has been found for the cosmological H {\sc ii} region, which is the fundamental prediction of today's reionization models.
Although there are many ingredients to be considered, we conclude that our conjectures about reionization are plausible and provide a reasonable explanation of our results.

\section{Conclusions}

We carried out spectroscopic observations with Subaru and Keck to identify z6p5LAEs that were selected by NB excess at $\sim920$ nm.
Our conclusions can be summarized as follows.

1. We have identified eight new z6p5LAEs based on their significantly asymmetric Ly$\alpha$ emission profiles. This increases the sample of spectroscopically confirmed $z=6.5$ LAEs in the SDF to $17$.

2. We have constructed a large, homogeneous spectroscopic sample from the photometric sample of $58$ LAE candidates to determine the Ly$\alpha$ LF at $z=6.5$. 
The LF reveals an apparent deficit, at least at the bright end, compared to that at $z=5.7$.
The $L^*$ difference between $z=6.5$ and $5.7$ is $\sim0.75$ mag for fixed $\alpha=-1.5$.
The difference has $3$ $\sigma$ significance, which decreases to $2$ $\sigma$ when we take into account cosmic variance.
There may also be a decrease in comoving number density of LAEs from $z=5.7$ to $6.5$ at the faint end, although this conclusion could be modified by further follow-up spectroscopy.

3. The rest-UV continuum LF of our LAE sample at $z=6.5$ is almost the same as that of the LAE sample at $z=5.7$ and the $i$-dropout sample at $z\sim6$, even at their bright ends.

4. The spatial distribution of z6p5LAEs was found to be homogeneous over the field, based on three independent methods to quantify the clustering strength.
We cannot rule out the possibility that we happen to see a very homogeneous region of the $z=6.5$ universe.

5. The composite spectrum of the $12$ z6p5LAE objects with high spectral resolution clearly reveals an asymmetric Ly$\alpha$ profile with an extended red wing.
The profile can be explained by either a galactic wind model composed of double Gaussian profiles or by a reionization model expected for a typical profile during the reionization epoch.

6. Our results could imply that the reionization of the universe has not been completed at $z=6.5$.
The decline of the z6p5LAE LF implies $0 \lesssim x_{\rm H I}^{\rm IGM} \lesssim 0.45$ based on the IGM dynamical model of \citet{san04}.
We conjecture that we are observing the final stage of reionization at $z=6.5$, when the LAE overdense regions were still obscured by surrounding thick, neutral IGM clouds, which qualitatively agrees with our results of deficient bright LAEs and their homogeneous spatial distribution.

Our z6p5LAE spectroscopic sample is not yet large enough to make a more precise comparison with LFs at lower redshifts.
At the moment, it is not clear whether the true LF of z6p5LAE is closer to our upper or lower limit estimates.
Thus, we do not conclude whether differences in $L^*$ or $\phi^*$ dominate in the disagreement of LF between $z=6.5$ and $5.7$.
In this study, we could not constrain $\alpha$, the faint-end slope of the LF.
The faint end of the z6p5LAE sample would critically determine the LAE contribution as ionizing sources to the photon budget of cosmic reionization.
Even if the number density of LAEs is low, a significant number of another star-forming population, LBGs, if they exist at this high-$z$ epoch, could complete the reionization.
Alternatively, only a small number of galaxies with huge star formation rates at very high redshift can reionize the universe \citep{pan05, mob05}.
It is too difficult to sample LBGs at exactly the same redshift as LAEs, whereas it is not certain what fraction of LBGs at this epoch shows strong enough Ly$\alpha$ emission to be observed as LAEs.
Understanding the evolutionary connection between LBGs and LAEs is linked to this problem.
\citet{shi06} found from the rest UV luminosity LF of the $z=5.7$ LAE sample that $\sim80\%$ or more of the LBG population would have strong Ly$\alpha$ emission at $z\sim6$.
This study shows that the rest-UV luminosity LF at $z=6.5$ agrees with that of $z=5.7$, suggesting the same high fraction.
The faint end of the LF of z6p5LAEs may barely be determined with the spectroscopic capability of today's $8$ m telescopes.
Nevertheless, steady efforts toward further spectroscopic confirmation are certainly required for z6p5LAEs.


\acknowledgments

We are grateful to the Keck Observatory staff for their help with the observation.
We deeply appreciate the devoted technical and management support of the Subaru Telescope staff for this long-term project.
The observing time for part of this project was committed to all the Subaru Telescope builders.
We thank an anonymous referee for helpful comments that improved the manuscript.
This research was supported by the Japan Society for the Promotion of Science through Grant-in-Aid for Scientific Research 16740118.

\appendix
\section{Skewness: the asymmetry indicator of high-$z$ Ly$\alpha$ emission lines}

We describe here a statistic {\it skewness} $S$ to measure the asymmetry of high-$z$ Ly$\alpha$ emission lines.
This model-independent indicator is free from fitting procedures that sometimes require smoothing of noisy spectra.

The expression $S$ is a popular statistic, defined as the third moment of a distribution function, which describes its asymmetry (see also \citealp{kur04}).
Here, we regard the observed spectrum, which is basically a two-dimensional array of the flux ($f_i$) and the pixel ($x_i$), as a distribution function with an array size of $n$.
Then $S$ is defined as
	
\begin{eqnarray}
S=\frac{1}{I\sigma^3}\sum^n_i(x_i-\overline{x})^3 f_i,
\end{eqnarray}

where $I=\sum^n_i f_i$, and $\overline{x}$ and $\sigma$ are the average and dispersion of $x_i$ defined as

\begin{eqnarray}
\overline{x}=\frac{1}{I}\sum^n_i x_i f_i, 
\end{eqnarray}
and
\begin{eqnarray}
\sigma^2=\frac{1}{I}\sum^n_i(x_i-\overline{x})^2 f_i, 
\end{eqnarray}
respectively.

This statistic is free from fitting procedures such as the $a_\lambda$ and $a_f$ presented by \citet{rho03}.
Their asymmetry estimation is based on two-component Gaussian profile fitting for the red and blue sides of the emission.
However, it is sometimes too difficult to determine accurately the peak wavelength $\lambda_p$ of the line, or $\lambda_{10}$, where the flux drops to $10\%$ of its peak value, since they strongly depend on the resolving power and quality of the data.

We now estimate the error in $S$, which can be analytically derived.
We here assume $\delta x_i\sim0$ and approximately regard the 1st-order derivative of skewness as its error,

\begin{eqnarray}
\delta S=\left(\sum^n_i \left(\frac{\partial S}{\partial f_i}\delta f_i \right)^2\right)^{0.5}=
\frac{1}{I}\left(\sum^n_i 
\left[ \left(\frac{x_i-\overline{x}}{\sigma}\right)^3 
- \frac{3S}{2} \left(\frac{x_i-\overline{x}}{\sigma}\right)^2 
- 3 \left(\frac{x_i-\overline{x}}{\sigma}\right)
+ \frac{S}{2}\right]^2\delta f_i^2\right)^{0.5},
\end{eqnarray}

where we use

\begin{eqnarray}
\frac{\partial \overline{x}}{\partial f_i}=\frac{1}{I}(x_i-\overline{x}), 
\end{eqnarray}

and

\begin{eqnarray}
\frac{\partial \sigma}{\partial f_i}=\frac{1}{2\sigma I}\left\{(x_i-\overline{x})^2-\sigma^2\right\}. 
\end{eqnarray}

We can assume that the flux error does not depend strongly on the wavelength, as in the case of our NB filter coverage, where the night-sky window is almost free from OH emission lines.
Under this assumption, the $\delta f_i$ can be regarded as $\delta f_i\sim\delta f=const.$ dominated by readout noise around the emission line in the spectrum.

This error estimate is confirmed to agree well with the rms fluctuation of $S$ evaluated by a Monte Carlo realization, in which the line-profile model was disturbed with random artificial errors as large as $\delta f$.
Our estimate of skewness error in equation (A4) gives a useful analytic formula, although a more strict error estimate can only be achieved directly with Monte Carlo simulations.

\epsscale{0.60}
\begin{figure}
\plotone{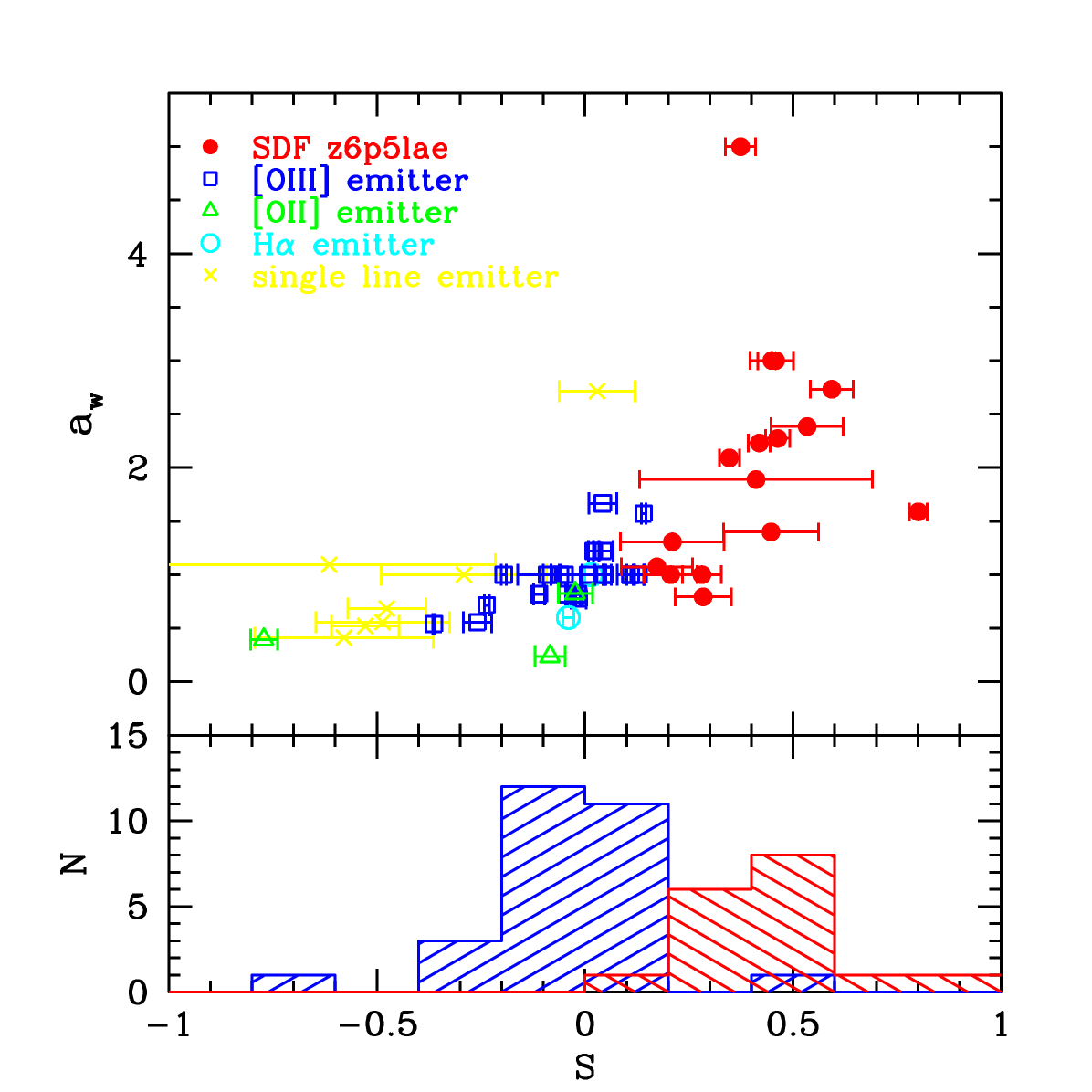}
\epsscale{1.0}
\caption{
{\it Top}: Comparison of the asymmetry indicator $S$ with $a_\lambda$ proposed by \citet{rho03}.
The filled circles denote objects classified as z6p5LAEs in this study, whereas squares, triangles, and open circles, are the apparent [O {\sc iii}], [O {\sc ii}], and H$\alpha$ emitters, respectively.
The crosses denote single-line emitters.
{\it Bottom}: The $S$ distribution for z6p5LAEs (right histogram) and foreground emitters (left histogram).
[{\it See the electronic edition of the Journal for a color version of this figure.}]
\label{fig_skew}}
\end{figure}

\epsscale{0.60}
\begin{figure}
\plotone{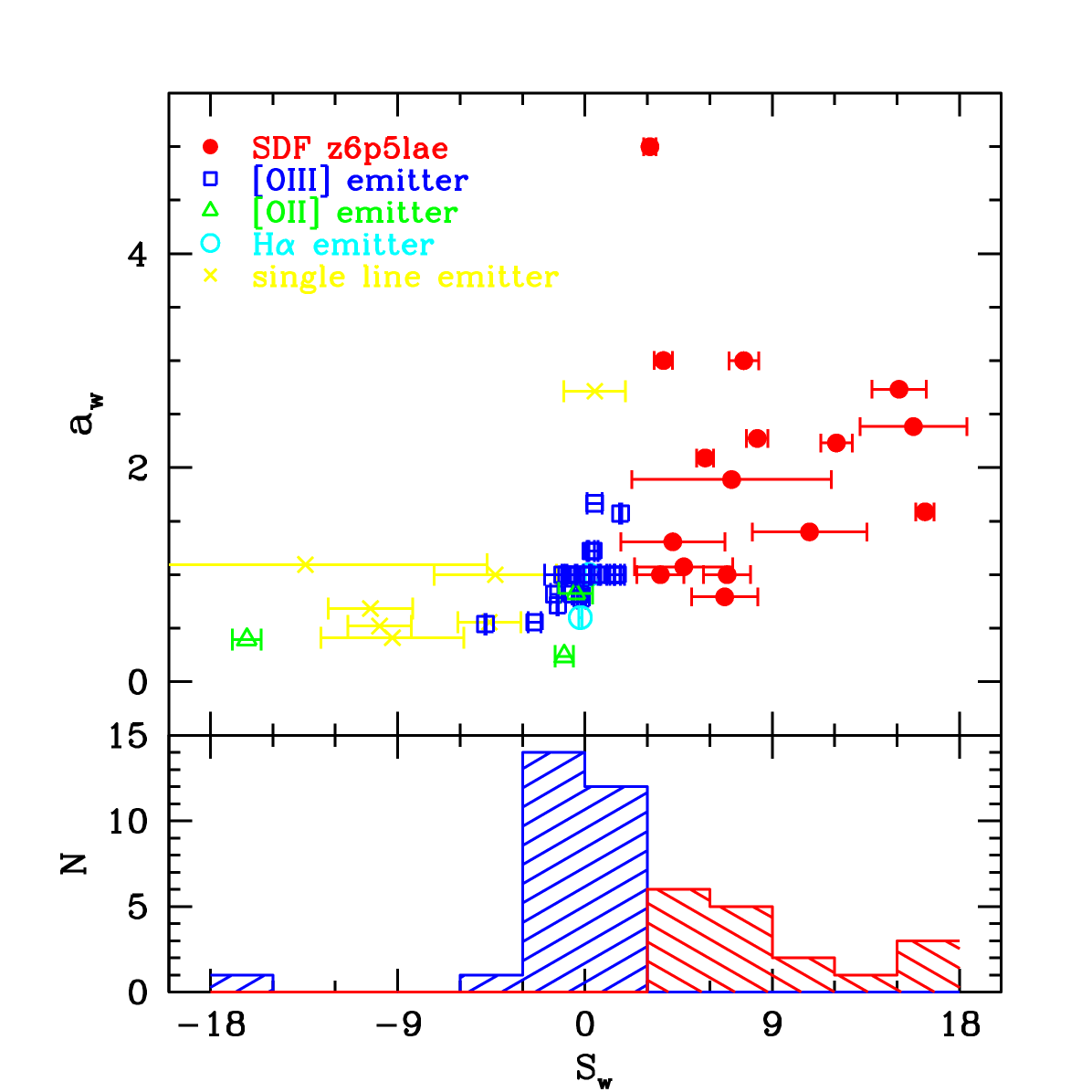}
\epsscale{1.0}
\caption{
Same as Fig.~\ref{fig_skew}, but for $S_w$.
The errors in $\lambda_{10,r}$ and $\lambda_{10,b}$ are not taken into account, and the error in $S_w$ is simply derived using $\delta S (\lambda_{10,r}-\lambda_{10,b})$.
[{\it See the electronic edition of the Journal for a color version of this figure.}]
\label{fig_rskew}}
\end{figure}

We compared this statistic $S$ with $a_\lambda$, proposed by \citet{rho03}, for our $NB921$-excess sample in Figure~\ref{fig_skew}.
We calculated $S$ in the effective wavelength range of an emission line, where $f_i$ has an apparent positive signal above the continuum (sky) level.
The slight change in the effective wavelength range does not significantly affect the result.
Foreground emitters shown in Figure~\ref{fig_skew} were definitely identified by their multiple spectral lines.
That is, an H$\alpha$ emitter has corresponding $[$O {\sc iii}$]$ doublets around $7016$ \AA, and $[$O {\sc iii}$]$ and [O {\sc ii}] emitters show apparent doubles by themselves (and sometimes H$\beta$ for the $[$O {\sc iii}$]$ case).
As expected, [O {\sc ii}], $[$O {\sc iii}$]$, and H$\alpha$ emitters are distributed around $S=0$ and $a_w=1$.
The resolved [O {\sc ii}] doublet lines are expected to show negative $S$ because $\lambda$3726 is typically weaker than $\lambda$3729.
There is a population (filled circles) that have actually larger positive $S$ and $a_w$, indicating statistically asymmetric lines with a broad red wing.
We recognize these as Ly$\alpha$ emitters that have $S>0.15$.
Almost all the single-line emitters (triangles) show relatively low $S$, which indicates that they are likely to be $[$O {\sc ii}$]$ emitters.
The large scatter of $S$ for these single-line emitters was caused by the low S/N of their spectra.

To find a further adequate indicator sensitive to an asymmetry, we introduce {\it weighted skewness} $S_w$, which combines the $S$ indicator and Rhoads et al. (2003)'s method.
High-$z$ Ly$\alpha$ emission usually has a wider FWHM in the observed frame than that of foreground emitters.
We define the {\it weighted skewness} $S_w$ as

\begin{eqnarray}
S_w=S(\lambda_{10,r}-\lambda_{10,b}), 
\end{eqnarray}

where $\lambda_{10,r}$ and $\lambda_{10,b}$ are the wavelengths where the flux drops to $10\%$ of its peak value at the redder and bluer sides of the Ly$\alpha$ emission, respectively.
The asymmetric index $S$ is a dimensionless quantity, whereas the $S_w$ has a dimension of wavelength (here measured in angstroms).
Figure~\ref{fig_rskew} shows the correlation between $S_w$ and $a_w$, where the symbols are the same as those in Figure~\ref{fig_skew}.
We found that all the apparent foreground emitters have $S_w<3$.
The $S_w$ can distinguish between Ly$\alpha$ and other lines more effectively than $S$.
We can set the critical value $S_w=3$ to distinguish LAEs and foreground emitters, although there could be more or less contamination of LAEs at $S_w<3$.
This is a conservative critical value for LAEs in the sense that it assumes low contamination and low completeness.

In summary, our proposed statistic $S$ is free from fitting procedures and has an analytical error estimation formula.
In addition, we made use of a revised indicator $S_w$ that proves to be more powerful when combined with the line width determined from Gaussian fitting.

\end{document}